\documentclass[11pt]{article}
	
	\usepackage[acronym,nonumberlist]{glossaries}
	\usepackage[english]{babel} 
	\usepackage[round]{natbib}
	\usepackage{a4wide}
	\usepackage{afterpage}
	\usepackage{amsmath,amsfonts,amsthm,amssymb}
	\usepackage{array} 
	\usepackage{bbm}
	\usepackage{bbold}
	\usepackage{booktabs}
	\usepackage[labelfont=bf,singlelinecheck=false]{caption}
	\usepackage{color}
	\usepackage{dsfont}
	\usepackage{enumerate}
	\usepackage{epstopdf}
	\usepackage{fancyhdr}
	\usepackage{float}
	\usepackage{floatflt}
	\usepackage{geometry}
	\usepackage{graphicx}
	\usepackage{hyperref}
	\usepackage{latexsym}
	\usepackage{lipsum}
	\usepackage{longtable}
	\usepackage{moreverb}     
	\usepackage{multicol}
	\usepackage{multirow}
	\usepackage{natbib}
	\usepackage{paralist}
	\usepackage{pdflscape}
	\usepackage{placeins} 
	\usepackage{ragged2e} 
	\usepackage{setspace}
	\usepackage{subcaption}
	\usepackage{tabularx}       
	\usepackage{titlesec}
	\usepackage{url}
	\usepackage{verbatim}

	\usepackage[table,xcdraw]{xcolor}
	\usepackage{colortbl}
	\usepackage{rotating}
	
	\usepackage{dcolumn}
	
	\hypersetup{
		colorlinks 	= true,
		linkcolor 	= blue,
		urlcolor 	= red,
		citecolor 	= blue
	}
	\newcolumntype{L}[1]{>{\raggedright\let\newline\\\arraybackslash\hspace{0pt}}m{#1}}
	\newcolumntype{C}[1]{>{\centering\let\newline\\\arraybackslash\hspace{0pt}}m{#1}}
	\newcolumntype{R}[1]{>{\raggedleft\let\newline\\\arraybackslash\hspace{0pt}}m{#1}}
	
	\definecolor{c2}{rgb}{0.6, 0.0, 0.1}

	\newcommand{\ltab}{\raggedright\arraybackslash} 
	\newcommand{\rtab}{\raggedleft\arraybackslash} 
	\newcommand{\Keywords}[1]{\par\noindent
		{\small{\em Keywords\/}: #1}}

    \newcommand{\hide}[1]{}

\begin{document}
	
\title{\Large The Role of Binance in Bitcoin Volatility Transmission } 

\vspace{20pt}

\author{\large Carol Alexander\footnote{University of Sussex Business School. Email: \href{mailto:c.alexander@sussex.ac.uk}{c.alexander@sussex.ac.uk}} $\,$, Daniel F. Heck\footnote{University of Sussex Business School. Email: \href{mailto:d.heck@sussex.ac.uk}{d.heck@sussex.ac.uk}} $\,$ and Andreas Kaeck\footnote{University of Sussex Business School. Email: \href{mailto:a.kaeck@sussex.ac.uk}{a.kaeck@sussex.ac.uk}}\\
}
%
%
\maketitle

\thispagestyle{empty}

\begin{abstract}
	\normalsize
	\noindent We analyse high-frequency realised volatility dynamics and spillovers between centralised crypto exchanges that offer spot and derivatives contracts for bitcoin against the U.S.  dollar or the stable coin tether. The tether-margined perpetual contract on Binance is clearly the main source of volatility, continuously transmitting strong flows to all other instruments and receiving very little volatility from other sources. We also find that crypto exchanges exhibit much higher interconnectedness when traditional Western stock markets are open. Especially during the U.S.  time zone, volatility outflows from Binance are much higher than at other times, and Bitcoin traders are more attentive and reactive to prevailing market conditions. Our results highlight that market regulators should pay more attention to the tether-margined derivatives products available on most self-regulated exchanges, most importantly on Binance.
\end{abstract}  

\bigskip



\thispagestyle{empty}

\newpage
\setcounter{page}{1}
\onehalfspacing
\doublespacing


\hide{Every asset class has its own unique market microstructure.} The microstructure of crypto asset and derivatives markets is very complex, very different from that of other asset classes, and not yet well understood in the academic literature. One key difference to traditional markets is its fragmentation, with numerous self-regulated, centralised and decentralised, exchanges operating in peripatetic tax-haven locations to avoid scrutiny of their practices and profits. On the other hand, the US-based Coinbase and other centralised exchanges like Bitstamp and Kraken now provide a relatively secure market place for trading spot crypto-crypto and crypto-fiat pairs. 

Currently, it is the self-regulated derivatives exchanges that are of most concern to regulators. \cite{alexander2020bitcoin} and other empirical research has established the dominance of these exchanges in price discovery of the major crypto assets, especially for bitcoin (BTC). The main aim of this paper is to study the high-frequency transmission of volatility from the self-regulated derivatives exchanges to the U.S.  or European regulated spot exchanges and to investigate potential time variation in the transmission mechanism throughout the trading day.

Understanding volatility flows is important for centralised exchanges and their market makers, who tend to operate simultaneously on several exchanges. A transmission of volatility into the exchange is a signal for market makers to widen spreads. This is because order slippage increases with realised volatility, and slippage reduces profits either directly, if the market maker pays the slippage costs, or indirectly because it induces a downwards pressure on order flow. To offset the slippage costs that are expected from an inflow of volatility, the market maker will try to increase spreads. Note that trading fees differ widely across exchanges.  Those with greater trading volume, like Coinbase and Binance, also tend to have higher realised volatility and a large, highly-competitive set of market makers. But responding to volatility flows by increasing spreads is difficult in this environment, as is passing on slippage costs to clients -- both would reduce the order flow. A reduction in order flow is also bad for exchanges, which seek to avoid a `toxic flow' situation. This is where there is a surfeit of  professional traders -- such as brokers, over-the-counter desks, high-frequency traders and the exchange's own designated market makers -- who prefer to trade with uninformed retail traders rather than other informed  traders. For this reason, exchanges with many market makers provide incentives for large trades by lowering their fees. 

Understanding volatility flows is also important for European and U.S. regulators who recently responded to the rapid growth in crypto trading during 2021 with extensive proposals for reform.\footnote{In July 2021 the European Commission proposed new regulations for Markets in Crypto Assets (MiCA) which makes cryptocurrency exchanges responsible for the loss of consumer assets because of fraud, cyber-attack or negligence. Shortly afterwards a U.S. senator Don Beyer presented congress with a 58-page draft of the Digital Asset Market Structure and Investor Protection Act,  proposing sweeping reforms and clarifying the responsibilities of the Federal Reserve, the SEC and the CFTC.} The proposals exemplify concerns about self-regulated exchanges such as Binance, which act not only as a trading platform, but also as broker, custodian and central counterparty for clearing trades. Knowing whether the main source of volatility on Coinbase, Kraken, Bitstamp and other regulated crypto exchanges comes from Binance, or other exchanges of concern, is very useful to inform the ongoing debates surrounding these regulatory reforms.

To the best of our knowledge, no previous study has analysed high-frequency volatility flows within and between the major crypto spot and derivatives exchanges and we aim to fill this gap. We analyse realised volatility spillovers through the lens of a multivariate version of the Logarithmic Multiplicative Error Model (LogMEM; \citet{Bauwens2000, Nguyen2020}). Formally, the model decomposes realised volatility into a product of its conditional mean and an error term which has unit mean and follows a distribution with non-negative support. The conditional mean is usually assumed to be autoregressive, depending both on its own lagged values and those of other variables of interest.\footnote{Since we analyse realised volatility, we also include an asymmetric response component to capture the leverage effect typically observed in financial markets \citep{Bollerslev2006}. LogMEMs are particularly well suited for modelling realised volatility dynamics as they ensure the non-negativity of volatility and can be extended to allow for occasional zero observations which can occur in high-frequency time series.} To estimate the model, we derive robust realised volatility measures from the spot BTC-USD price on three spot exchanges (Coinbase, Kraken and Bitstamp) and three self-regulated exchanges (Binance, and Huobi) and from the prices of the main bitcoin perpetual contracts traded on Binance and another self-regulated exchange, Bybit. Contrary to the existing literature, we base our analysis on high-frequency data, using second-by-second observations to compute realised volatility at the five-minute frequency which, due to microstructure noise, is the highest frequency which allows a reliable transmission analysis. All other studies on volatility spillovers between different cryptocurrencies \citep{Yi2018, Katsiampa2019a, Wang2020, Caporale2021, Sensoy2021} or between individual crypto-exchanges \citep{Cheah2018, Ji2021} rely on daily observations, and/or they focus exclusively on the much smaller spot market, which could lead to erroneous conclusions. By analysing realised volatility on major spot and derivatives exchanges at the five-minute frequency we study more informative volatility flows at a much more granular level.

\begin{figure}[tb]
	\centering
	\caption{Volatility Flows Between the Main BTC-USD Instruments}
	\includegraphics[width=0.7\textwidth]{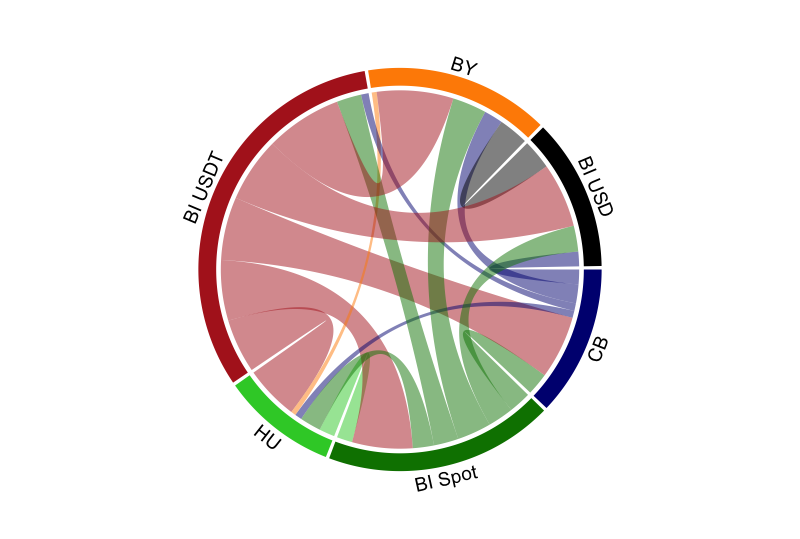}
	\vspace{-3ex}
	{\raggedright \justify \footnotesize \textit{Note:} The figure shows the magnitude of volatility flows between six major BTC-USD instruments including: Coinbase spot (CB); Binance spot (BI Spot); Huobi spot (HU); Binance tether-margined BTC perpetual (BI USDT), Binance USD perpetual (BI USD) and Bybit USD perpetual (BY). The flows are estimated using a multivariate LogMEM on the five-minute realised volatility between 1 January and 31 March 2021. The parameters  that are not significant are set to zero. The direction of flow is indicated by its colour, and the size is proportional to the width at its origin or destination, these being equal. 
	Our findings suggest that trading bitcoin against tether on Binance is the main source of volatility. Binance's indirect USD perpetual emits much less volatility and mainly receives flows from the tether-margined product. \par}
	\label{fig:flow_plot}
\end{figure}

Our main empirical finding is that  the tether-margined perpetual contract on Binance is the main emitter of volatility, continuously transmitting strong flows to all other instruments throughout the day. And, out of all instruments included in our analysis, this contract also receives the lowest volatility flows. Figure \ref{fig:flow_plot} depicts the overall magnitude of the volatility flows detected by our LogMEM as a circular plot, omitting Bitstamp and Kraken because prior analysis shows Coinbase to be the main volatility transmitter within spot markets. The Binance tether-margined perpetual exhibits very strong, positive volatility spillovers to all other instruments, while it receives only weak (and negative) volatility flows from Binance spot, even less from Coinbase, and no significant transmission from any other instrument.   Both Coinbase and Binance spot also have a negative volatility transmission to the other instruments, and the Binance USD perpetual only transmits volatility to the Bybit perpetual. Volatility shocks on the perpetual swaps are very short lived, so we can explain these flows if traders on spot exchanges need longer to react to these shocks that traders on perpetuals. That is, once the volatility has risen on  spot exchanges, the transient volatility increase in perpetuals has already been reversed. Finally, both Bybit and Huobi are receivers rather than transmitters of volatility. Both their fee structure and their time-of-day volume patterns indicate that Bybit and Huobi attract different types of traders to Binance. Because these and the spot exchanges generate much weaker volatility flows than Binance, they have substantially less contagion potential. 

Next we study whether the strength of the Binance volatility outflows varies within the trading day. We find that they are lowest during Asian trading,  intensify substantially during European trading and finally reach their maximum during U.S.  trading hours. Not only the outward volatility flows but also their short-term persistence strengthens over the course of the day. Thus, when updating their expectations, traders are more attentive and reactive to prevailing market conditions during U.S.  trading hours. The bitcoin market also exhibits a higher interconnectedness when traditional Western stock markets are open. 
Taken together with a truly remarkable periodicity observed in  time-of-day volume patterns, our results demonstrate that Western (especially U.S. based) traders are using the highly-controversial Binance exchange to trade bitcoin.

A further contribution of our paper is to highlight the important role of the stable coin tether, which is pegged 1:1 to the U.S. dollar. Besides a few recent studies such as \citet{Griffin2020, Ji2021, Baur2021},  tether has not attracted significant attention from academics or regulators. Yet,  tether has become central to the entire crypto market, helping to ensure sufficient liquidity and activity. At the time of writing tether has a market capitalisation of around \$62bn, up from \$10bn in July 2020. On most days, its trading volume far exceeds any other crypto asset. For example, on 30 June 2021 tether's trading volume was almost \$63bn, exceeding that of bitcoin (\$34bn) and ether (\$26bn) combined.  

We make several other contributions to the literature. First, we reveal a remarkable intra-day pattern of both trading volume and realised volatility on crypto instruments that to the best of our knowledge has not yet been documented in the related literature. Between midnight and 16:00 UTC, both volume and volatility evolve in a clear U-shape. After that, all instruments exhibit a continuous decrease in volume and volatility to an average level. The minimum is generally attained during early UTC morning. In fact, the intraday pattern found on crypto instruments seems similar to traditional FX markets, such as the Deutsche Mark/USD exchange rate \citep{Andersen1997}. Strikingly, we also find distinct four-hourly spikes in both trading volume and realised volatility that are very pronounced on the first five minutes of every eighth hour (i.e. from 00:00 to 00:05, 08:00 to 08:05 and 16:00 to 16:05 UTC). We relate these unusual spikes to the funding payments on the perpetuals. Possibly, some market participants modify their positions across multiple exchanges just after funding on some perpetual has occurred or they take advantage of some mispricing between spot and perpetuals.  Secondly, since realised volatility and trading volume are deeply interconnected and usually go hand-in-hand, one might ask if results on volatility spillovers are mirrored by trading volume flows.  To answer this question, we repeat the LogMEM analyses replacing realised volatility by five-minute trading volume. The volume flow results are in sharp contrast to the volatility transmissions. Volume flows mainly from the spot instruments, especially on Coinbase and Binance, but also from the Bybit perpetual. There are negligible volume outflows from the main emitter of volatility, i.e. Binance's tether-margined perpetual.

Finally, we shed more light on the relationship between trading volume and realised volatility. This highly-relevant microstructure question has been studied extensively in traditional asset classes. For example, \citet{Nguyen2020} find that trading volume positively affects volatility, and  volatility negatively affects trading volume. Similarly, \citet{Chevallier2012} document a positive and statistically significant effect of trading volume on realised volatility for both oil and gas futures. However, within the crypto market, the volume-volatility relationship has attracted only little attention. To the best of our knowledge, no study has yet examined this issue using high frequency data on multiple instruments. We document a strong significant bi-directional causality relationship between trading volume and realised volatility at the five-minute frequency. Using a bivariate LogMEM, we find that an increase in volatility is preceded by an increase in trading volume in the previous five-minute interval. However, in the subsequent five-minute period, the relationship is reversed, leading to a negative effect of volume on volatility and a positive influence of volatility on volume. We attribute this inversion to a longer response time (i.e. a time delay) of volatility.

The remainder of this paper is organised as follows: 
Section \ref{sec:exch} describes the exchanges and instruments included in our analysis; Section \ref{sec:data} describes our second-by-second data, outlines the realised volatility calculation and discusses the characteristics of intra-day patterns of 5-minute volume and realised volatility; Section \ref{sec:meth} explains the econometric framework; 
Section \ref{sec:results} presents our results on volatility transmission, both within and between different instrument groups, and provides an economic interpretation of the intra-day variation; Section \ref{sec:volume} examines volume flows between the different instruments, and the volume-volatility relationship; and Section \ref{sec:concl} summarises and concludes.

\section{Exchanges and Instruments}\label{sec:exch}

Trading on crypto assets and their derivatives has grown substantially since July 2020, when a landmark ruling by the Office of the Comptroller of the Currency permitted their custody  in federally chartered U.S.  banks. From \$275 trillion in July 2020, the total market capitalization of crypto assets rose to a peak of over \$1,750 trillion in May 2021, on a monthly trading volume exceeding \$5 trillion. The notional traded on the major crypto derivatives exchanges grew even faster over this period, from less than \$500 bn in July 2020 to around \$6 trillion in May 2021.\footnote{Sources: \href{https://www.statista.com/statistics/730876/cryptocurrency-maket-value/}{Statistica} and \href{https://www.cryptocompare.com}{CryptoCompare Exchange Review} }

Despite continuous growth, the crypto spot market remains tiny compared with the market for derivatives which remains self-regulated, with the notable exception of the bitcoin and ether futures and options on the Chicago Mercantile Exchange (CME). Most self-regulated exchanges offer two types of bitcoin derivatives products, namely inverse (also called coin-margined) and linear contracts. The former are rather complicated products: for ease of use, they are quoted in USD but their actual quote currency is bitcoin so they are contracts on USD/BTC instead of BTC/USD. This way, the self-regulated exchanges can avoid onboarding fiat currencies entirely, even though crypto-fiat prices are traded. On the other hand, the linear products are exactly like derivatives in traditional asset classes -- i.e. they are denominated, margined and settled in the same currency. However they do not use a fiat currency and instead use a digital currency such as tether (USDT) or another stable coin.


We only admit the major (well-established) bitcoin spot exchanges (Bitstamp, Coinbase, Huobi, Kraken) as well as major (self-regulated) crypto derivatives exchanges into our analysis (Binance, Bybit). However, we exclude futures from our analysis and focus on the very popular perpetual futures, also called perpetual swaps, or simply perpetual contracts, which generally exhibit much higher trading volumes. Table \ref{tab:perp_specs} reports the product specifications for different perpetual contracts. The USD-contracts are of inverse type, i.e. their base currency is BTC so they are contracts on USD/BTC instead of BTC/USD. The tether-contracts however are ordinary linear products, that is they are settled in tether. As is usual with crypto products, the perpetuals allow very high leverage and can be traded 24/7.\footnote{ These contracts have most in common with currency swaps  (either crypto--crypto or crypto--fiat) where the only cash-flows are between perpetual fixed and floating legs and these are usually exchanged every 8 hours via a funding payment that is designed to tie the perpetual contracts price to the spot price. Thus, the funding rate is positive when the perpetuals price exceeds the spot, but can be negative when the spot exceeds the perpetuals price. Please note that even though the crypto--fiat contracts are usually denoted as BTCUSD for example -- in traditional FX markets, this quotation would mean that BTC is the base currency and USD is the quote currency -- all cashflows are in cryptocurrencies so that exchanges do not have to deal with traditional fiat currencies and their onboarding process.
But they also combine the features of both futures and spot positions in that they do not expire before being closed out (similar to a spot position) and they allow very high-leverage trading (like a futures contract). In contrast to ordinary futures, perpetual contracts are not exposed to any roll-over risk and the basis is very much smaller than it is for futures.} 

\begin{table}[tb]
	\centering
	\caption{Perpetuals Specifications}
	\vspace{-2ex}
	\footnotesize
	\begin{tabular}{p{4.45cm}p{3.6cm}p{3.6cm}p{3.6cm}}
		\midrule
		& \multicolumn{2}{c}{USD Contracts} & \multicolumn{1}{c}{USDT Contracts}\\
		\cmidrule(r){2-3} \cmidrule(r){4-4}
		& Binance & Bybit & Binance \\
		\midrule
		Type & Inverse & Inverse & Linear \\
		Contract Size  & 100 USD & 1 USD & 0.001 BTC \\
		Margin Requirement  & 0.8\%*   & 1\%   & 0.8\%* \\
		Settlement Currency  & BTC   & BTC   & USDT   \\
		Trading Days  & 24/7  & 24/7  & 24/7  \\
		Funding Frequency & 8 hrs & 8 hrs & 8 hrs  \\
		Fees (maker/taker) & 1/5  & -2.5/7.5 & 2/4 \\
		Tick Size & 0.1 USD & 0.5 USD & 0.01 USDT\\
		\midrule
	\end{tabular}%
	\vspace{0.5ex}
	\footnotesize
	{\raggedright \justify \textit{Note:} The table shows the main specifications of the perpetual contracts included in our analysis. All fees reported here are in basis points. *The leverage on Binance depends on the notional value of the position. The larger the position, the lower the leverage allowed. \par}
	\label{tab:perp_specs}
\end{table}

A key point to note from Table \ref{tab:perp_specs} is that the fees on all the self-regulated derivatives exchanges follow (at least partly) a maker-taker model. That is, orders that add liquidity to the book (non-marketable limit orders) are charged less than orders that reduce liquidity (market orders or marketable limit orders). In the case of Bybit, maker fees are even negative, i.e. liquidity-increasing orders are refunded a certain percentage of the order size, independent of the user's past trading volume. For their USD perpetual, Binance follow a tiered maker-taker model refusing ordinary users maker rebates. Only VIP investors with a 30-day trading volume of at least 50,000 BTC and a balance of 2,000 of their in-house token (called Binance Coin; BNB) obtain maker rebates. 
Similarly, the spot exchanges do not offer maker rebates and follow a volume-dependent fees schedule of at most 50bps (Coinbase and Bitstamp), 26bps (Kraken), 20bps (Huobi) and 10bps (Binance). Bitstamp is the only spot exchange not offering discounts for liquidity-increasing orders.

Binance, Bybit and Huobi are still completely self-regulated. There is no supervisory authority establishing any rules to prevent malpractice, misconduct and manipulation. 
Admittedly, it is a very challenging task to regulate these venues because the traders could simply switch to some other venue still offering very loose or no regulation, of which there are plenty. However, even though we cannot rule out the possibilitys of price or volume manipulation, this does not prevent a proper spillover analysis. If anything, the lack of supervision makes our study more interesting for market participants and regulators.

\section{Prices, Volumes and Realised Volatility}\label{sec:data}
\begin{figure}[tb]
	\centering
	\caption{Bitcoin Price from 1 January to 31 March 2021}
	\vspace{-1ex}
	\includegraphics[width=\textwidth]{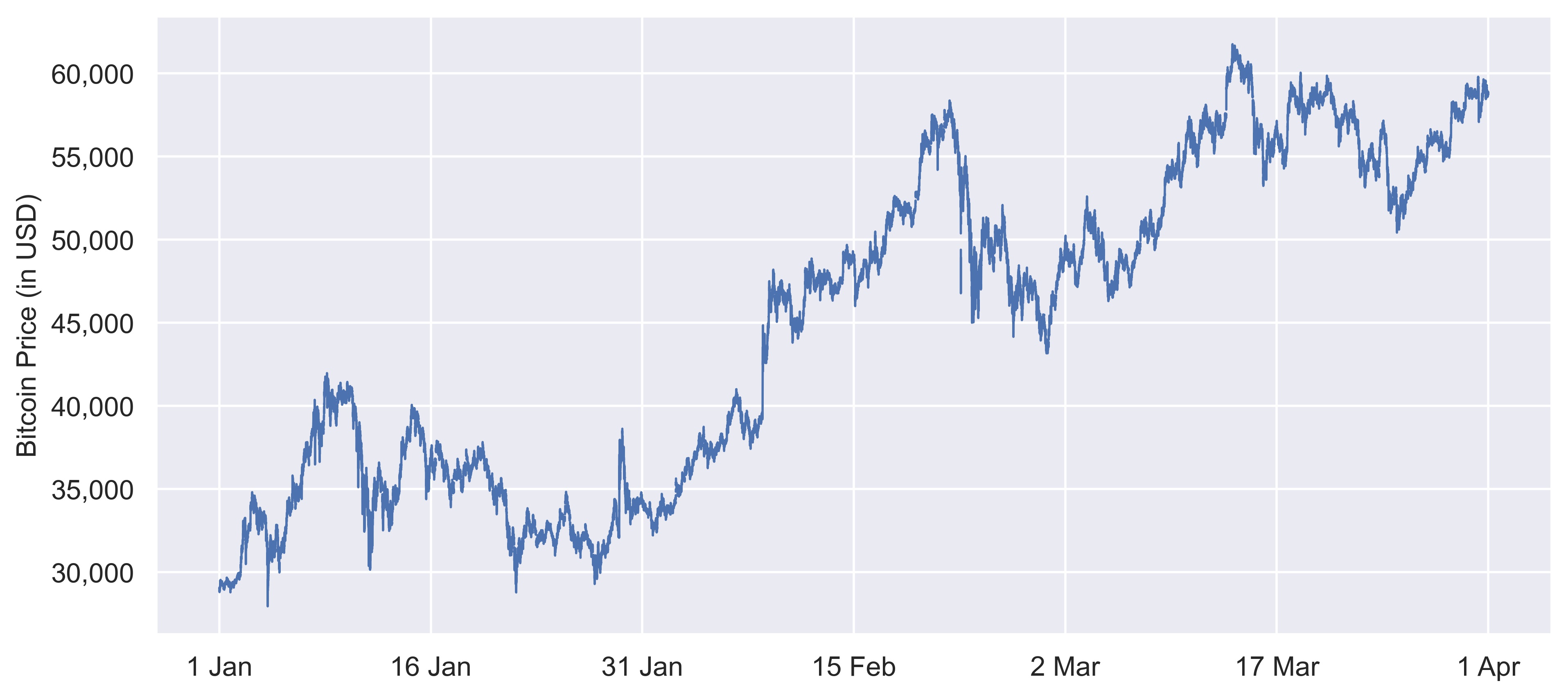}
	\vspace{-3ex}
	{\raggedright \justify \footnotesize \textit{Note:} The figure shows the bitcoin price on Coinbase in USD, from 1 January to 31 March 2021. \par}
	\label{fig:XBT_price}
\end{figure}

The crypto market evolves at an extremely fast pace -- not only through innovations such as Decentralised Finance (DeFi) and Non-Fungible Tokens (NFTs) but also due to actions of regulators in the US, China and, more recently, many countries particularly concerned about Binance's activities. Therefore,  we are interested in high-frequency inter-exchange volatility flows, rather than any long-term dynamics. For this reason we choose a one-second data sampling frequency so that we are able to compute realised volatility every five minutes. We focus on the recent bull market from 1 January to 31 March 2021, where the price of bitcoin rose by almost 100\% from \$30,000 to around \$60,000. Figure \ref{fig:XBT_price} shows the detailed price evolution over this three-months period.

We retrieved data on all instruments from \href{https://www.coinapi.io/}{coinAPI}.\footnote{Data and software provider coinAPI links with hundreds of crypto spot and derivatives exchanges, offering historical and streaming order-book and trades in tick-by-tick or highest granularity data from all major centralised and decentralised exchanges.} The dataset consists of second-by-second price and volume data and covers the period from 1 January to 31 March 2021, summing up to more than 7.5 million observations per instrument and almost 60 million in total. Table \ref{tab:data} provides summary statistics on the trading activity of the different instruments included in our analysis. We can see that trading on perpetuals (both USD and USDT) and BTC/USDT spot pairs is highly active. Transactions occur in at least three out of four seconds -- for the tether-instruments, this fraction is even more than 95\% -- and the average daily volume (ADV) exceeds \$1.5 bn. The highest ADV occurs on the Binance USDT perpetual with even more than \$15 bn. Out of the three USD spot pairs, only Coinbase is able to keep up. Transactions are conducted in 84\% of one-second intervals and its ADV is about \$1.2 bn. With trading volumes of less than half a billion U.S.  dollar and transactions occurring only in about every fourth second, Bitstamp and Kraken lag far behind.

\begin{table}[tb]
	\centering
	\small
	\caption{Volume Data Statistics}
	\vspace{-2ex}
	    \begin{tabular}{p{2cm}p{2cm}p{2cm}>{\rtab}p{3.5cm}>{\rtab}p{2.5cm}>{\rtab}p{2.5cm}}
	    	\midrule
	    	\multicolumn{1}{l}{Type} & \multicolumn{1}{l}{Currency} & Instrument & \rtab Count & \multicolumn{1}{r}{ADV}  & \multicolumn{1}{r}{MaxDV} \\
	    	\midrule
	    	\multicolumn{1}{l}{Spot} & USD & Bitstamp & 1,867,238 (24\%) &  392.92  & 1,364.60 \\
	    	& & Coinbase &  6,554,500 (84\%) &       1,187.42  &      3,499.38 \\
	    	& & Kraken & 1,777,536 (23\%) &           345.03  &           893.02  \\
	    	\cmidrule(r){2-6}
	    	& USDT & Binance  &  7,716,626 (99\%) &           3,858.29  &    8,410.34        \\
	    	& & Huobi &  7,420,413 (95\%) &       1,586.62      &      4,194.10      \\
	    	\midrule
	    	\multicolumn{1}{l}{Perpetuals} & USD & Binance & 5,798,339 (75\%) &  5,481.04   &  15,084.64    \\
	    	& & Bybit & 6,748,841 (87\%) &   7,368.11   &   17,755.14   \\
	    	\cmidrule(r){2-6}
	    	& USDT & Binance & 7,712,699 (99\%) &    15,006.25  &    36,297.44  \\
			\midrule
	\end{tabular}%
	\vspace{0.5ex}
	{\raggedright \justify \footnotesize \textit{Note:} The table shows the number of second intervals where at least one trade was conducted (Count, in absolute terms and as percentage of the total number of second-intervals), the average daily volume (ADV, in million USD) and the maximum daily volume (MaxDV, in million USD), during the period from 1 January to 31 March 2021. \par}
	\label{tab:data}%
\end{table}%

\begin{figure}[p]
	\centering
	\small
	\caption{Volume Intra-Day Pattern}
	\vspace{-1ex}
	\includegraphics[width=\textwidth]{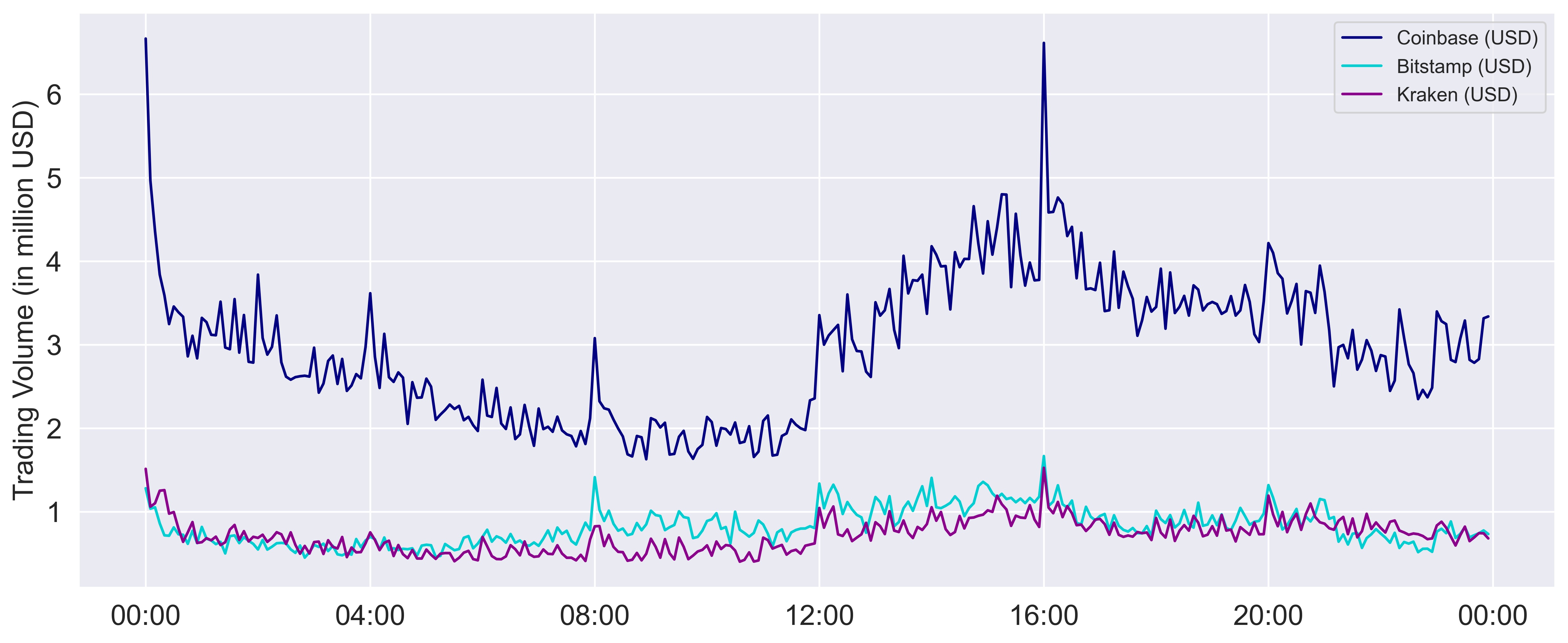}
	\vspace{0.5cm}
	\includegraphics[width=\textwidth]{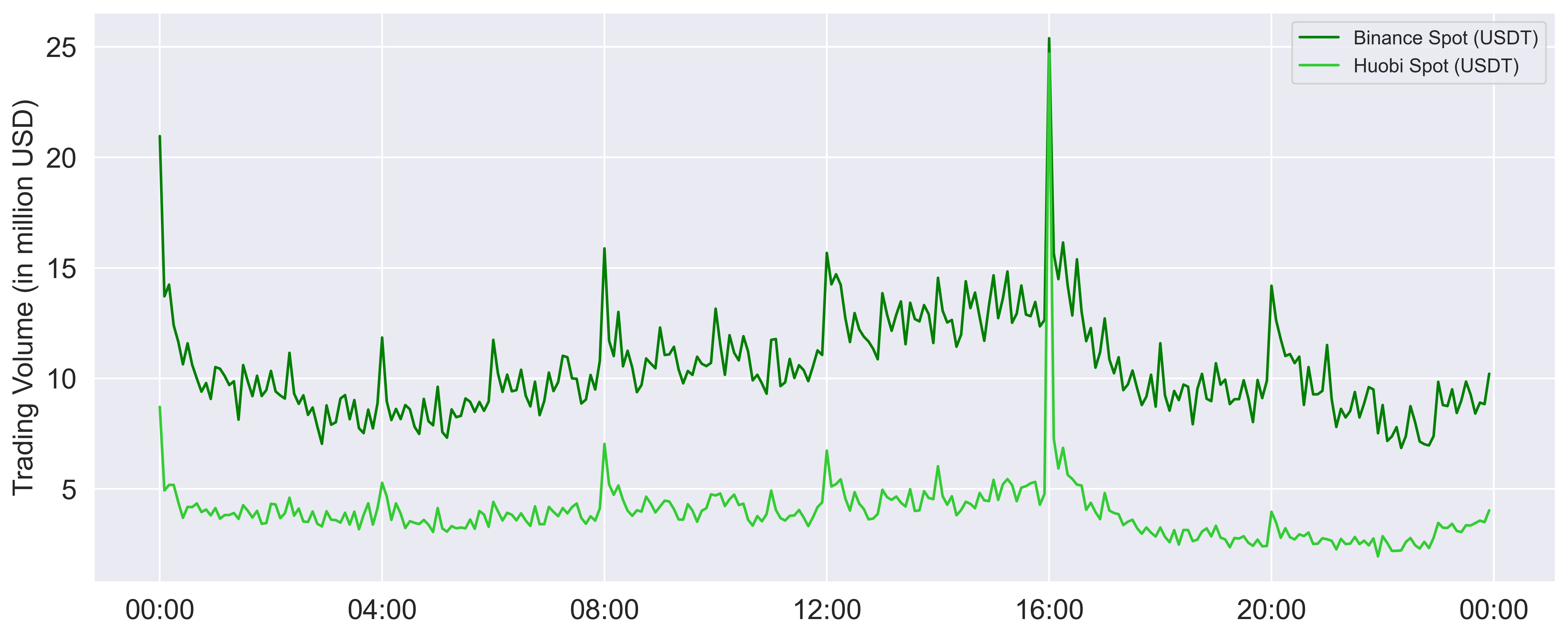}
	\vspace{0.5cm}
	\includegraphics[width=\textwidth]{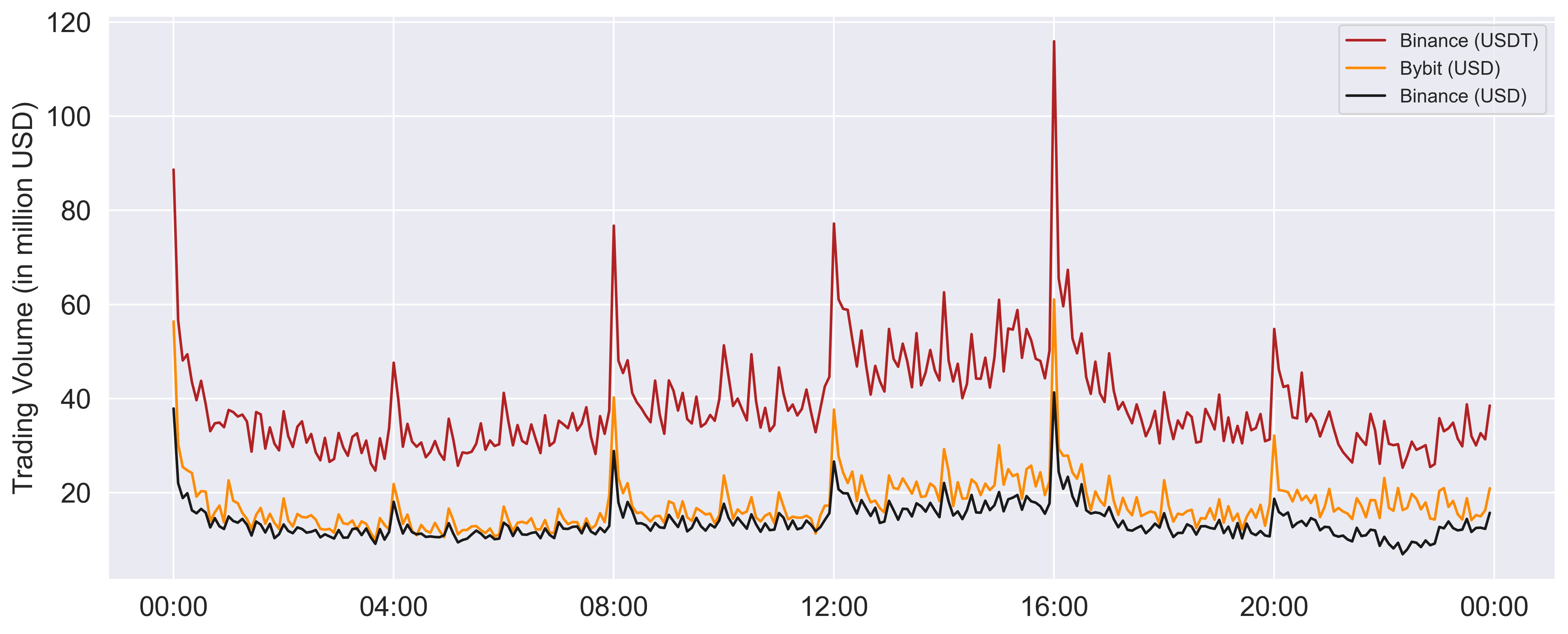}
	\vspace{-6ex}
	{\raggedright \justify \footnotesize \textit{Note:} The figure shows the intra-day pattern of five-minute trading volume (in million USD) for USD spot pairs (upper graph), USDT spot pairs (middle graph) and perpetuals (lower graph), measured as the median five-minute volume over the period from 1 January to 31 March 2021. All times are in UTC. \par}
	\label{fig:intra-day_volume}
\end{figure}

It is well documented that trading volume exhibits a certain intra-day pattern. In North American equity markets, many studies find a U-shaped pattern \citep{Jain1988,McInish1990}, while \citet{Cai2004} and \citet{Ozenbas2008} document a more M-shaped behaviour on the London Stock Exchange. For major currency pairs, \citet{Danielsson2001} and \citet{McGroarty2009} find a similar M-shaped volume pattern with peaks at London and New York opening times. Figure \ref{fig:intra-day_volume} shows the intra-day pattern of trading volume for BTC/USD and BTC/USDT, both on spot exchanges and perpetual contracts, measured as the median five-minute trading volume over the period from 1 January to 31 March 2021.\footnote{Note that the number of transactions shows an analogous intra-day pattern.} As can be seen, volume follows a similar pattern on spot and perpetuals. First, it evolves in a U-shape with very distinct peaks at midnight and 16:00 UTC. The most extreme spike can be observed on the Huobi USDT spot pair, where the median trading volume increases fivefold from around \$5 m to almost \$25 m at 16:00 UTC. After the afternoon spike, trading volume on all instruments continuously decreases to an average level. The three USD spot pairs reach their minimum between 09:00 and 10:00 UTC, while the volumes on perpetuals and USDT spot pairs seem to be at their lowest in the early morning around 03:00 UTC. Consequently, we can only partly confirm the results of \citet{Eross2019} who document an inverted U-shape on Bitstamp with a peak at around 14:00 UTC. However, their data spans  2014 to 2017 and so the difference in intra-day trading volume could be explained by a significant evolution of the crypto market over the last three years.

The two graphs of Figure \ref{fig:intra-day_volume} show another interesting feature. Trading volume exhibits distinct spikes that are very pronounced on the first five minutes of every eighth hour, i.e. from 00:00 to 00:05, 08:00 to 08:05 and from 16:00 to 16:05 UTC. 
There is only one explanation for these spikes that we can think of, namely the funding payments on the perpetuals.\footnote{Some spikes are evident every 4 hours. While the contracts considered here (Binance, Bybit) are funded at 00:00, 08:00 and 16:00 UTC, other major perpetuals such as the BitMEX one exchange cash flows at 04:00, 12:00 and 20:00 UTC.} Possibly, some market participants hold accounts across multiple spot and derivatives exchanges and systematically modify their positions just at the time of funding payments. This could be  to take advantage of some mispricing between spot and perpetuals. 

Next we analyse time-of-day patterns in realised volatility. Using our second-by-second price data for any given exchange, we calculate its realised volatility at a five-minute frequency. This seems to be a reasonable trade-off between too much microstructure noise at a higher frequency and potentially missing significant volatility flows at lower frequencies \citep{Andersen2000, Nguyen2020}. 
Since there is still a lot of noise in the data, we calculate realised volatility using the robust estimator based on pre-averaging \citep{Jacod2009}. Here, the one-second log returns are locally smoothed using a weighted average, which reduces the influence of microstructure noise to some extent. In particular, the pre-averaged log returns are given by 
\begin{equation*}
\overline{r}_t = \sum_{j=1}^{k_n} g\left(\frac{j}{k_n}\right) r_{t+j}
\end{equation*}
where $r_t$ denotes the ordinary (not pre-averaged) log returns and $g$ is a real-valued weighting function, commonly chosen as $g(x) = \min (x,1-x)$.\footnote{The only requirements on the weighting function $g:[ 0, 1 ] \rightarrow \mathds{R}$ are that it is continuous, piecewise continuously differentiable with piecewise Lipschitz derivative, and having $g(0)=g(1)=0$. Since the tent-shaped function $g(x) = \min (x,1-x)$ is the simplest functions meeting these criteria, it is commonly chosen as weighting function.} Following the empirical analysis of \citet{Hautsch2013}, the bandwidth $k_n$ of the local pre-averaging window is chosen to be $\lceil \theta \sqrt{n} \, \rceil$ where $n$ is the number of observations (in our case, $n=300$) and $ \theta \in \left[  0.3, 0.6 \right]$. The exact choice of the smoothing parameter $\theta$ depends mostly on the sampling frequency and the liquidity of the asset. Since we are interested in accuracy for  multivariate spillovers, rather than individual volatility estimates, we set $\theta$ consistently as 0.4 for all instruments. Therefore, the optimal bandwidth is $\lceil 0.4 \sqrt{300} \, \rceil = 7$. For each five-minute interval, the pre-averaging estimator of realised variance is then calculated as the sum of squared pre-averaged one-second returns. Finally, to obtain the pre-averaged realised volatility, we take the square root of the realised variance and annualise it by a factor of $\sqrt{12 \times 24 \times 365}$.\footnote{This pre-averaging reduces the influence of zero returns on the realised volatility measure. If no transaction has been executed within a one-second interval, we fill this gap with the previous price which leads to a zero return and therefore induces a downward bias in realised volatility. Even without pre-averaging, using the last available price is the most appropriate way to deal with one-second intervals in which no transaction has taken place, because it seems intuitive that an instrument with lower trading activity should also exhibit a lower realised volatility. }


Since the crypto market is still not as liquid as more established asset classes, it is possible that  a very limited number of transactions occur during some five-minute intervals. Therefore, we apply a threshold on trading activity in order to obtain reliable volatility estimates. If there are transactions in less than 20\% of the one-second intervals, we set the volatility estimate of that five-minute period equal to zero. Given the volume data statistics in Table \ref{tab:data}, this threshold will be of little or no concern for Coinbase, Huobi and the three Binance instruments, but it may be relevant for Bitstamp and Kraken. For this reason, we use a  zero-augmented log-normal distribution in the LogMEM to account for zero observations in the conditional mean specification explicitly.  This way, it will not become a major issue if these two exchanges occasionally exhibit a comparatively large number of zero values.

Table \ref{tab:realvola_stats} reports summary statistics for five-minute realised volatility over the period from 1 January to 31 March 2021. The average levels of volatility on Coinbase, Bitstamp, Huobi, Bybit and the three Binance instruments are similar, but that of Kraken is significantly lower, probably due to comparatively low trading activity. For the same reason, Bitstamp and Kraken show rather a high standard deviation of more than 60\% -- compared to about 43\% on the six remaining instruments -- paired with a zero 25\%-quantile. We also note that all eight instruments show a minimum volatility of zero, implying that our trading activity threshold is not met at least once for all instruments. Overall, except for Bitstamp and Kraken, the realised volatility statistics are quite similar. Finally, it is striking that Kraken shows an extremely high maximum volatility of 63. This outlier value was attained on 22 February 2021 between 14:20 and 14:25 UTC 
when the price on Kraken jumped up and down with an increasing amplitude 
resulting in a very high realised volatility not reflected by the other exchanges. This extremely large bid-ask bounce was probably caused by some exchange-specific issues, such as server problems. To reduce, but not entirely eliminate, the impact of such anomalous values we winsorize the realised volatility data before our empirical analysis.

\begin{table}[tb]
	\centering
	\small
	\caption{Realised Volatility Statistics}
	\begin{tabular}{c|rrr|rr|rrr}
		& \multicolumn{1}{c}{\textbf{Coinbase}} & \multicolumn{1}{c}{\textbf{Bitstamp}} & \multicolumn{1}{c|}{\textbf{Kraken}} & \multicolumn{1}{c}{\textbf{Binance$^{S}$}} & \multicolumn{1}{c|}{\textbf{Huobi}} & \multicolumn{1}{c}{\textbf{Binance$^{T}$}} & \multicolumn{1}{c}{\textbf{Bybit}} & \multicolumn{1}{c}{\textbf{Binance$^{\$}$}} \\
		\hline
		\textbf{Mean} & 0.6985 & 0.6217 & 0.4709 & 0.7237 & 0.7299 & 0.7630 & 0.6146 & 0.7068 \\
		\textbf{Std} & 0.4255 & 0.6102 & 0.6837 & 0.4155 & 0.4147 & 0.4371 & 0.4408 & 0.4409 \\
		\textbf{Min} & 0     & 0     & 0     & 0     & 0     & 0     & 0     & 0 \\
		\textbf{25\%} & 0.4344 & 0     & 0     & 0.4650 & 0.4721 & 0.4909 & 0.3347 & 0.4311 \\
		\textbf{50\%} & 0.5945 & 0.6262 & 0.4012 & 0.6235 & 0.6302 & 0.6534 & 0.5044 & 0.5990 \\
		\textbf{75\%} & 0.8326 & 0.9504 & 0.7358 & 0.8541 & 0.8620 & 0.8972 & 0.7562 & 0.8464 \\
		\textbf{Max} & 11.6779 & 8.0250 & 63.3657 & 8.5440 & 7.8419 & 7.4741 & 6.6263 & 7.7105 \\
		\midrule
	\end{tabular}
	{\raggedright \justify \footnotesize \textit{Note:} The table shows summary statistics on the five-minute realised volatility over the period from 1 January to 31 March 2021. Binance$^{S}$, Binance$^{T}$ and Binance$^{\$}$ denote the BTC/USD spot pair, the USDT perpetual and the USD perpetual, respectively, on Binance.\par}
	\label{tab:realvola_stats}
\end{table}

\begin{figure}[tb]
	\centering
	\small
	\caption{Volatility Intra-Day Pattern}
	\vspace{-1ex}
	\includegraphics[width=\textwidth]{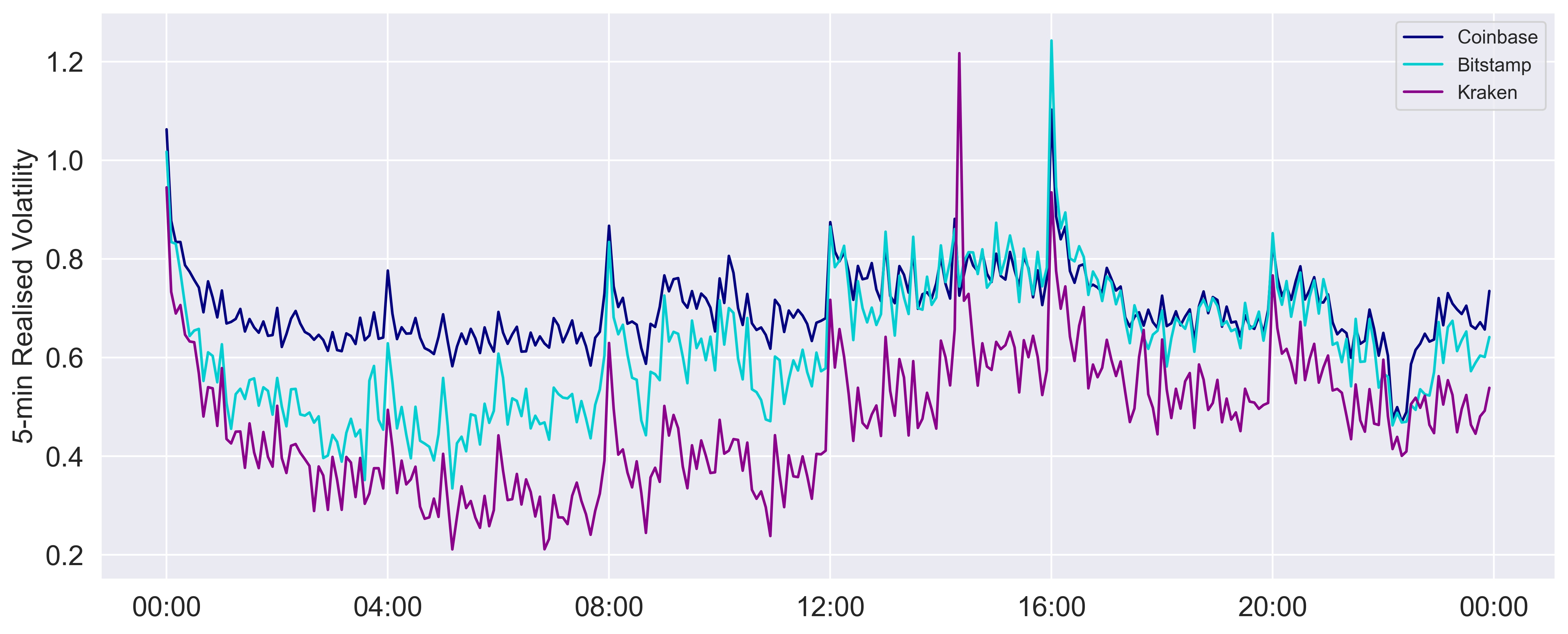}
	\vspace{0.5cm}
	\includegraphics[width=\textwidth]{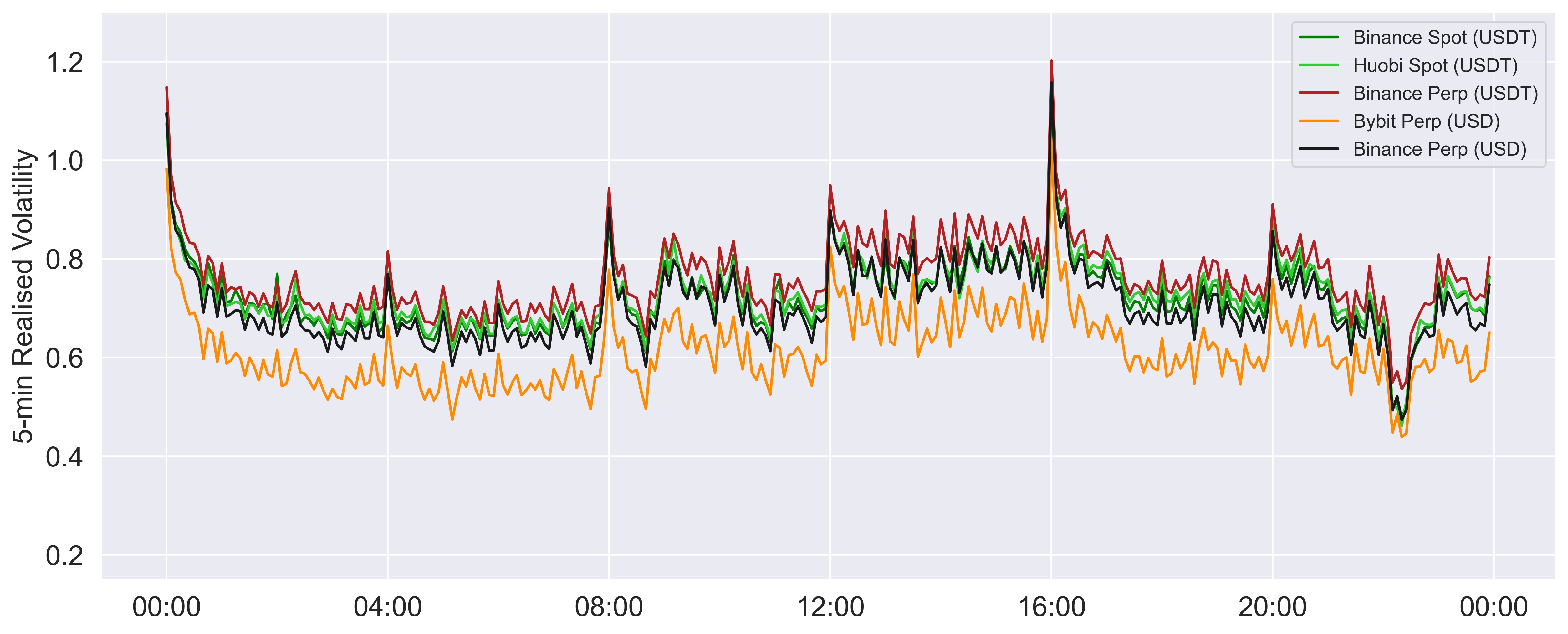}
	\vspace{-6ex}
	{\raggedright \justify \footnotesize \textit{Note:} The figure shows the intra-day pattern of 5-minute realised volatility (in million USD) for USD spot pairs (upper graph) as well as USDT spot pairs and perpetuals (lower graph), measured as the average five-minute realised volatility over the period from 1 January to 31 March 2021. All times are in UTC. \par}
	\label{fig:intra-day_volatility}
\end{figure}

The intra-day pattern of realised volatility is closely related to that of volume \citep{Berger2009,Bubak2011}.  Figure \ref{fig:intra-day_volatility} depicts the five-minute realised volatility on the  spot exchanges (above) and the other, self-regulated exchanges below.  As with trading volume it first follows a U-shape with distinct peaks at midnight and 16:00 UTC and smaller spikes at 04:00 and 08:00 UTC. After 16:00, it slowly declines to between 50\% and 80\%, with another visible spike at 20:00 UTC. In the upper graph, we can also see that Bitstamp and Kraken exhibit a lower volatility than Coinbase and the remaining USDT spot pairs and perpetual contracts. The difference is especially pronounced from midnight until 11:00 UTC, i.e. during Asian and early European trading hours. Therefore, we suppose that the active traders on Bitstamp and Kraken are mainly American and European. 
Otherwise the eight instruments exhibit a  similar intra-day pattern for realised volatility. Finally, we note that the extremely large volatility on Kraken around 14:20 UTC is caused by the exchange-specific outlier discussed above and will be winsorized at 0.05\% in our empirical analysis. 

Apart from the spikes at times of funding payments on the perpetuals, the intra-day pattern of bitcoin volatility seems similar to the results of \citet{Andersen1997} for the Deutsche Mark/USD exchange rate. They find that volatility starts at a relatively high level around 00:00 UTC, then slowly decays for about 3 hours. After that, trading activity increases to a local maximum around the opening time of European markets. The global maximum  is attained at the opening of U.S.  markets. Afterwards, the volatility slowly declines and only starts to pick up again around 8 hours later.

It is important to account for this deterministic intra-day pattern in our empirical analysis, but we cannot  model the intra-day pattern explicitly in the MEM because  the number of parameters to be estimated would become too large. Therefore we do so in a non-parametric way by dividing each realised volatility observation by the average realised volatility of the respective five-minute interval.\footnote{To reduce the impact of extreme outliers, we could use the median five-minute volatility as a diurnal adjustment factor. However, Kraken exhibits zero median values in the early UTC morning and for this reason we must rely on the mean five-minute volatility. Apart from Kraken in the early morning, mean and median intra-day patterns have a very similar shape anyway.}  \citet{Nguyen2020} use a 250-day moving average as their diurnal adjustment factor but this is not possible in our study, so we simply use the average over the complete period from 1 January to 31 March 2021,  as depicted in Figure \ref{fig:intra-day_volatility}. Since our sample period is much shorter than in \citet{Nguyen2020}, time-variation is not of great concern. Then, as mentioned above, we Winsorize the top 0.05\% of our diurnally-adjusted realised volatility values, i.e. all observations greater than the 99.95\%-quantile are set equal to this quantile. 

\begin{figure}[tb]
	\centering
	\caption{Histograms of Five-Minute Realised Volatility}
	\vspace{-1ex}
	\includegraphics[width=\textwidth]{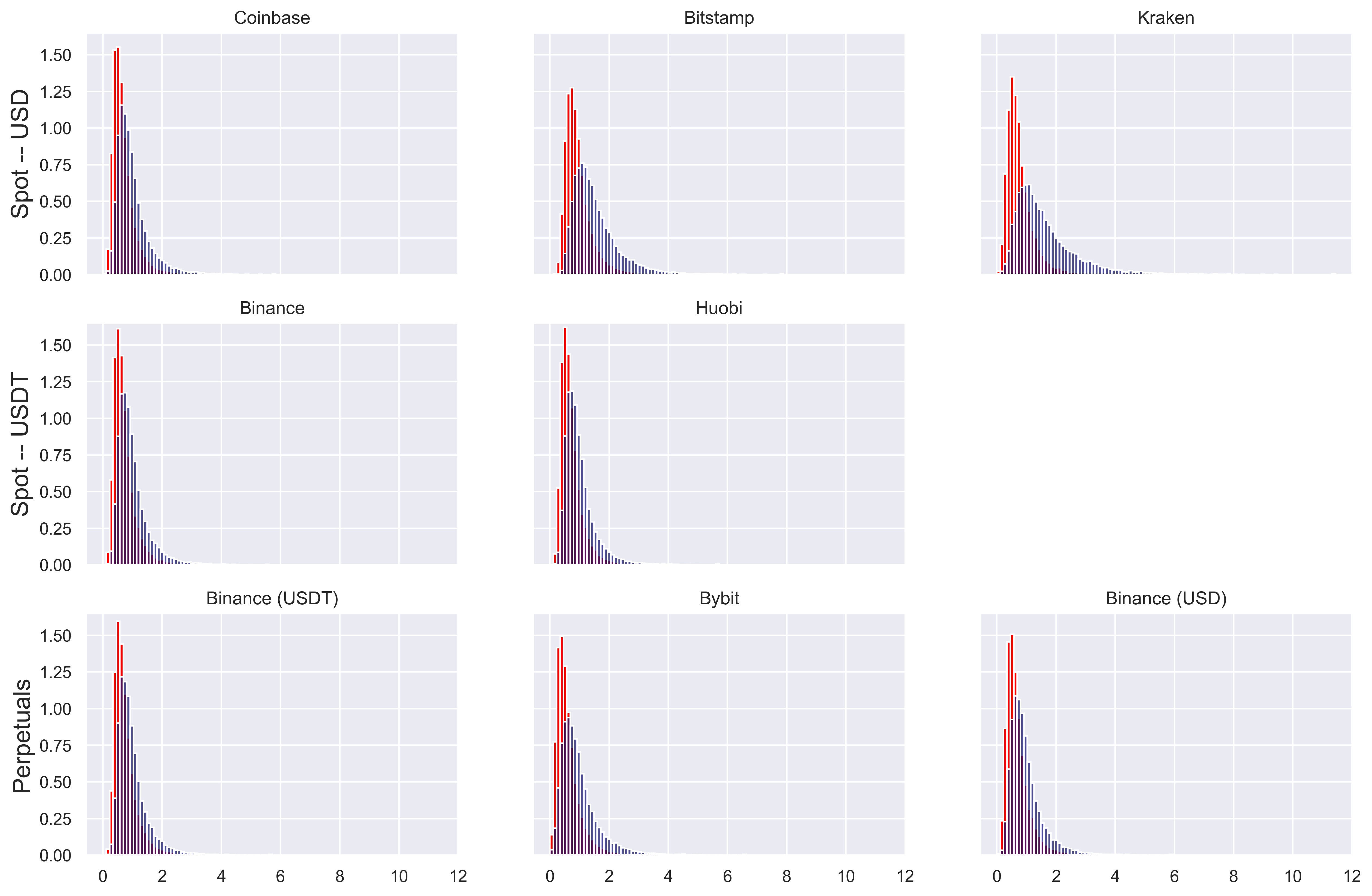}
	\vspace{-3ex}
	{\raggedright \justify \footnotesize \textit{Note:} The figure shows histograms of the five-minute realised volatility for USD spot pairs (top row), USDT spot pairs (middle row) and perpetuals (bottom row), both before (red) and after (blue) diurnal adjustment and Winsorizing, covering the period from 1 January to 31 March 2021. Note that due to the quite large number of zero observations on Kraken and Bitstamp, we excluded these zero values, allowing a better comparison across instruments. \par}
	\label{fig:histogram_grid}
\end{figure}

To see the effect of this data filtering, Figure \ref{fig:histogram_grid} exhibits the distributions of the five-minute realised volatility across the eight instruments, both before and after diurnal adjustment, winsorizing and excluding the relatively large number of zero observations for Kraken and Bitstamp, to allow a better comparison across instruments. These histograms confirm the summary statistics and graphs already discussed. That is, the realised volatility histograms for Coinbase, Huobi, Bybit and the three Binance instruments exhibit similar, low-range features. 
But due to relatively low trading activity, Bitstamp and Kraken exhibit quite a high portion of zero values and consequently, their histograms have lower modal values than those of the remaining six instruments. Overall, the histograms seem appropriate and we can use the filtered data in our empirical study.

\section{Methodology: Multiplicative Error Model}\label{sec:meth}
Since volatility is a non-negative variable, we base our analysis on the Multiplicative Error Model (MEM) introduced in \citet{Engle1998}.\footnote{Originally, the MEM has been proposed to model durations between transactions and it is therefore often called Autoregressive Conditional Duration (ACD) model. However, throughout this paper, we refer to the model and its extensions by the more general term Multiplicative Error Model.} This model implicitly guarantees the non-negativity of the variables of interest and overcomes certain difficulties and inefficiencies of the standard Gaussian approach when modelling non-negative time series.\footnote{As argued by \citet{Engle2002}, the two approaches that are usually applied when dealing with non-negative financial time series (either ignoring the non-negativity or taking logs) suffer severe disadvantages, like a very difficult maximum likelihood estimation or the need to work with very peculiar distributions.} 

In the original MEM, parameter restrictions have to be imposed to ensure non-negativity which exacerbates model estimation and interpretation. Therefore, we use the LogMEM(p,q) introduced by \citet{Bauwens2000}, where the logarithm is applied to the conditional mean in its specification, making any non-negativity constraints obsolete. Our univariate model for realised volatility $x$ is therefore given by
\begin{align}
		x_t & = \mu_{t} \, \varepsilon_t \label{eq:uniLogMEM1} \\
		\log \mu_{t} & = \omega + \sum_{j=1}^{p} \alpha_j \log x_{t-j} + \gamma \log x_{t-1}^- + \sum_{j=1}^{q} \beta_j \log \mu_{t-j} \label{eq:uniLogMEM1_condMeanSpec}
\end{align}
where $\varepsilon_t\geq0$ is a unit mean {iid} innovation, $\log x_{t-1}^{-} = \log x_{t-1}$ if the return of the respective interval is negative and zero otherwise, $p$ and $q$ determine the lag  structure and $(\omega, \alpha_1,\dots, \alpha_p, \gamma, \beta_1,\dots, \beta_q)$ are constant model parameters. As required, exogenous variables such as dummies to capture time variation may also be included in the conditional mean equation.

Finally, we need to specify the conditional distribution of the innovations. As shown by \citet{Allen2008}, the log-normal distribution yields consistency and asymptotic normality of the Quasi-Maximum Likelihood (QML) estimator in the LogMEM and exhibits superior finite sample properties compared to other widely-used error distributions. Since we analyse high-frequency time series, it is quite likely to encounter zero observations, which the log-normal distribution is not able to capture. Therefore, we follow \citet{Nguyen2020} and apply the zero-augmented version of the log-normal distribution as proposed by \citet{Hautsch2014}. Its density function is given by
\begin{equation*}
	f(x) = (1-p^+) \delta(x) + p^+ \frac{1}{\sqrt{2 \pi} \, x \, s} \exp \left( -\frac{1}{2} \left( \frac{\log(x) - m}{s} \right)^2 \right) \mathds{1}_{\{ x>0 \}}
\end{equation*}
where  $\delta(x)$ is a point probability mass at zero, $m = - 0.5 s^2 - \log p^+$ due to the requirement of unit mean and $p^+$ denotes the probability of strictly positive observations in the time series.\footnote{To limit the number of parameters in the LogMEM, we set $p^+$ equal to the empirical ratio of strictly positive observations, rather than estimating it when fitting the model.} Since the logarithm is only defined for strictly positive values, we modify the conditional mean specification in \eqref{eq:uniLogMEM1} by adding auxiliary parameters $\alpha^0$ which capture the effect of zero volatilites on the conditional mean, i.e.
\begin{equation}\label{eq:uniLogMEM1_zero}
	\log \mu_{t} = \omega + \sum_{j=1}^{p} \alpha_j \log x_{t-j} \, \mathds{1}_{\{ x_{t-j} > 0 \}}  + \sum_{j=1}^{p} \alpha_j^{0} \, \mathds{1}_{\{ x_{t-j} = 0 \}} + \gamma \log x_{t-1}^- + \sum_{j=1}^{q} \beta_j \log \mu_{t-j}.
\end{equation}

This specification can easily be extended to multiple dimensions to obtain the vector Logarithmic Multiplicative Error Model (vLogMEM(p,q)). It is given by
\begin{equation}\label{eq:vLogMEM1}
	\begin{split}
		\boldsymbol{x}_t &= \boldsymbol{\mu}_t \odot \boldsymbol{\varepsilon}_t  \\
		\log \boldsymbol{\mu}_t &= \textbf{w} + \sum_{j=1}^{p} \textbf{A}_j  \left( \log \boldsymbol{x}_{t-j} \odot \mathds{1}_{\{ \boldsymbol{x}_{t-j}>0 \}} \right)+ \sum_{j=1}^{p}  \textbf{A}_j^0 \mathds{1}_{\{ \boldsymbol{x}_{t-j}=0 \}} + \boldsymbol{\Gamma} \log \boldsymbol{x}_{t-1}^- + \sum_{j=1}^{q}  \textbf{B}_j \log \boldsymbol{\mu}_{t-j}
	\end{split}
\end{equation}
where $\odot$ denotes the Hadamard (element-by-element) product and the indicator functions should be interpreted component-wise. Model parameters are now collected in the vector \textbf{w} and the matrices $\textbf{A}_1,\dots,  \textbf{A}_p$, $\textbf{A}_1^0, \dots, \textbf{A}_p^0$, $\boldsymbol{\Gamma}$ and $\textbf{B}_1, \dots, \textbf{B}_q$.

The specification of the error distribution is much more challenging in the multivariate model, since the variables of interest are likely to be interdependent. There are three ways of dealing with this. First, we could use the multivariate log-normal distribution, which is one of the very few multivariate distributions with non-negative support \citep{Taylor2017}. This distribution is, however, not able to capture probability mass at zero and an appropriate adaption as in the univariate case is far from trivial. Second, we could apply copulae to link the univariate distributions of variables \citep{Cipollini2017}. Due to the non-trivial fraction of valid zero observations, this approach is quite challenging since we would have to estimate the zero-augmented mixture distributions and the copula jointly \citep{Nguyen2020}. Therefore, we follow the third option where the innovations of the individual variables are assumed to be orthogonal and only lagged interdependence is allowed. This assumption requires both the covariance matrix of innovations and the long-term persistence matrix \textbf{B} to be diagonal, which makes this approach equivalent to fitting the above univariate models for each variable individually, with the lagged values of the remaining instruments as exogenous variables. Even though this equation-by-equation approach leads to a loss in efficiency, it still yields consistent estimates in a QML context. Moreover, compared to a model allowing full interdependence, the procedure limits the number of parameters to be estimated and thus sidesteps the curse of dimensionality \citep{Escribano2018}.

\section{Volatility Transmission}\label{sec:results}

\subsection{Univariate Dynamics}\label{subsec:univariate_results}
To detect any differences between the instruments we fit the univariate LogMEM given in Equations \eqref{eq:uniLogMEM1} and \eqref{eq:uniLogMEM1_zero} to the diurnally-adjusted five-minute realised volatility of each instrument. We only report empirical results for models with one lag ($p=q=1$). As argued by \citet{Nguyen2020}, the maximum number of lags is expected to be three but it is lower in practice due to finite sample issues, for example.\footnote{\label{footnote_lags} As a robustness check, we have conducted a thorough analysis on the number of lags included in the LogMEM. In general, the total persistence -- the sum of all $\alpha$ and $\beta$ -- as well as the distribution parameter $s$ are independent of $p$ and $q$. The mean squared error and the Ljung-Box test statistic change only very marginally, while log-likelihood and BIC improve only when increasing the number of lags from one to two. The main difference seems to be the ``spread'' of the persistence across the $\alpha$ and $\beta$ parameters, which is in line with \citet{Nguyen2020}.} Table \ref{tab:uniMEM_results} reports parameter estimates, together with the log-likelihood value, the Bayesian Information Criterion (BIC) and the half-life of the conditional mean. Overall, the volatility dynamics are quite similar across all eight instruments. Not only is the total persistence of realised volatility -- the sum of $\alpha$ and $\beta$ -- very high, varying between 0.96 and 0.98 for all instruments, but also the individual degrees of short-term persistence (captured by $\alpha$) and long-term persistence (captured by $\beta$) are very similar at around 0.38$-$0.4 and 0.57$-$0.6, respectively. Only the Bybit contract stands out -- its short-term and long-term persistence are 0.29 and 0.68, respectively, which indicates that traders on Bybit are less reactive to prevailing market conditions than on the remaining seven instruments. A possible explanation might be related to the  trading activity resulting from Bybit's fee structure, which offers maker rebates of 2.5bps for all users, and the leverage 100 offered on  its perpetual contract which has a size of only \$1. These properties make the product very attractive for smaller retail traders who are usually considered rather uninformed and therefore, the disclosure of new information or a change in market conditions may have less influence on Bybit than on the other seven instruments.

\begin{table}[tb]
	\centering
	\footnotesize
	\caption{Univariate LogMEM(1,1)}
	\begin{tabular}{p{0.5cm}|>{\ltab}p{1.5cm}>{\ltab}p{1.5cm}>{\ltab}p{1.5cm}|>{\ltab}p{1.5cm}>{\ltab}p{1.5cm}|>{\ltab}p{1.5cm}>{\ltab}p{1.5cm}>{\ltab}p{1.5cm}}
	& \multicolumn{1}{c}{\textbf{Coinbase}} & \multicolumn{1}{c}{\textbf{Bitstamp}} & \multicolumn{1}{c|}{\textbf{Kraken}} & \multicolumn{1}{c}{\textbf{Binance$^{S}$}} & \multicolumn{1}{c|}{\textbf{Huobi}} & \multicolumn{1}{c}{\textbf{Binance$^{T}$}} & \multicolumn{1}{c}{\textbf{Bybit}} & \multicolumn{1}{c}{\textbf{Binance$^{\$}$}} \\
	\midrule
	$\omega$ & $-$0.0042$^{**}$ & $-$0.1675$^{***}$ & $-$0.1701$^{***}$ & $-$0.0069$^{***}$ & $-$0.0073$^{***}$ & $-$0.0056$^{***}$ & 0.0025 & $-$0.0042$^{**}$ \\
	$\alpha$   & 0.3741$^{***}$ & 0.3837$^{***}$ & 0.3808$^{***}$ & 0.3896$^{***}$ & 0.3859$^{***}$ & 0.4012$^{***}$ & 0.2871$^{***}$ & 0.3781$^{***}$ \\
	$\alpha^0$ & 0.0089 & $-$0.0648$^{***}$ & $-$0.2879$^{***}$ & $-$0.1964$^{*}$ & $-$0.0681 & 0.0020 & 0.0316 & $-$0.2373$^{***}$ \\
	$\gamma$& 0.0324$^{***}$ & 0.0379$^{***}$ & 0.0527$^{***}$ & 0.0326$^{***}$ & 0.0330$^{***}$ & 0.0305$^{***}$ & 0.0359$^{***}$ & 0.0345$^{***}$ \\
	$\beta$    & 0.5901$^{***}$ & 0.5995$^{***}$ & 0.5972$^{***}$ & 0.5733$^{***}$ & 0.5791$^{***}$ & 0.5610$^{***}$ & 0.6829$^{***}$ & 0.5851$^{***}$ \\
	$s$ 		   & 0.2837$^{***}$ & 0.2826$^{***}$ & 0.4825$^{***}$ & 0.2617$^{***}$ & 0.2545$^{***}$ & 0.2548$^{***}$ & 0.4074$^{***}$ & 0.2958$^{***}$ \\
	\hline
	LL & $-$753 & $-$21,217 & $-$25,898 & 503 & 1,314 & 1,448 & $-$8,353 & $-$1,594 \\
	BIC & 1,568 & 42,494 & 51,857 & $-$944 & $-$2,566 & $-$2,835 & 16,769 & 3,248\\
	\hline 
	h & \hphantom{$-$1,1}95 & \hphantom{$-$1,1}204 & \hphantom{$-$1,1}156 & \hphantom{$-$1}92 & \hphantom{$-$1,1}97 & \hphantom{$-$1,1}90 & \hphantom{$-$1,1}114 & \hphantom{$-$1,1}93 \\
	\midrule
\end{tabular}
	{\raggedright \footnotesize \justify \textit{Note:} The table reports parameter estimates, log-likelihood (LL), Bayesian Information Criterion (BIC) and half-life (h, in minutes) for the univariate LogMEM(1,1), fitted to diurnally-adjusted five-minute realised volatility over the period from 1 January to 31 March 2021. Binance$^{S}$, Binance$^{T}$ and Binance$^{\$}$ represent the BTC/USD spot pair, the USDT perpetual and the USD perpetual, respectively, on Binance, while the parameter $s$ denotes the standard deviations of the log residuals. The asterisks $^{***}$,$^{**}$, $^{*}$ indicate significance at the 1\%, 5\%, and 10\% level, respectively, based on robust standard errors. The half-life is calculated as $\log(0.5)/\log(\alpha+\beta)$, assuming strictly positive values and no asymmetric response.\par}
	\label{tab:uniMEM_results}
\end{table}


Both the highest short-term persistence and the lowest long-term persistence are observed on the Binance USDT perpetual. Interestingly, the two tether spot pairs (Binance, Huobi) exhibit the second and third highest (lowest) short-term (long-term) persistence. With standard errors of less than 0.01, the difference to the estimates for USD-instruments (Coinbase, Bitstamp, Kraken, Binance USD perpetual) is significant. Therefore, we conclude that when updating their expectations, traders on the tether-margined products -- especially on the Binance USDT perpetual -- pay more attention to current market conditions than those on the USD-margined instruments.

The half-life, which in our model measures the time it takes the logarithm of the conditional mean to halve its distance to the long-term mean, is quite similar for Coinbase, Huobi and the three Binance instruments.\footnote{The half-life is calculated as $\log(0.5)/\log(\alpha+\beta)$, thus assuming only strictly positive realised volatility values and no asymmetric response.} On these exchanges, it takes on average between 90 and 97 minutes until the effect of a volatility shock on the traders' expectation for the volatility within the next five-minute interval has halved. The lowest half-life is found on the Binance USDT perpetual, which once more indicates that the traders on this product pay most attention to the current state of the market when updating their expectations. On Bitstamp, Kraken and Bybit, it takes the traders' expectation much longer to revert to the long-term average, which is probably caused by quite low trading volume on Bitstamp and Kraken and rather uninformed trading activity on Bybit.

As expected, the estimate of the auxiliary parameter $\alpha^0$ is either not statistically significant or negative, therefore reducing the conditional mean after a zero observation occurs. Moreover, the estimate of the asymmetric response component is positive and highly significant on all eight instruments, which confirms the presence of the leverage effect often documented in other asset classes. The strength of this effect is very similar across instruments at around 0.03, only Kraken stands out with a comparatively high estimate of 0.05. This implies that a negative return increases the volatility expected by traders for the next five-minute interval by about 7\% to 10\% more than a positive return. Finally, the distribution parameter $s$ is quite similar on Coinbase, Bitstamp, Huobi and the three Binance instruments. Only Kraken and Bybit once again exhibit a significantly higher parameter estimate, indicating a greater standard deviation of innovations.

Compared to the findings of \citet{Nguyen2020} for the U.S.  Treasury market, all eight crypto instruments exhibit a much higher (lower) degree of short-term (long-term) persistence, implying higher sensitivity to prevailing market conditions and shorter memory than the Treasury notes. The total persistence however is higher on the U.S.  Treasury market, albeit only slightly (0.99 compared to between 0.96 and 0.98). Interestingly, during the 2007-2009 financial crisis as well as before and after economic announcements, the Treasury market exhibits elevated short-term persistence and its realised volatility dynamics appear to be more similar to those of crypto instruments.

\subsection{Multivariate Dynamics}\label{subsec:multivariate_results}
In this section, we analyse multivariate volatility flows among instruments. To provide an overview without substantially increasing the number of model parameters, we first fit the multivariate LogMEM given in Equation \eqref{eq:vLogMEM1} to each group of instruments separately. That is, we estimate the model for the three USD spot pairs (Coinbase, Bitstamp, Kraken), then we fit it to the two tether spot pairs (Binance, Huobi) and after that to the three perpetual swaps (Binance USDT, Bybit, Binance USD). Finally, we fit the model to all instruments jointly, except for Bitstamp and Kraken. Due to the comparatively large number of zero values, their exclusion makes estimation and interpretation of the model more straightforward.
To be consistent with the univariate analysis above, we estimate the models with one lag ($p=q=1$).\footnote{If more lags are included, the overall results behave very similar to the univariate case and do not change significantly (see footnote \ref{footnote_lags}).} Compared to the univariate analysis, the estimates for the intercept, the asymmetric response component and the distribution parameter do not change significantly and are thus not reported. Similarly, to economise on space we do not include the estimates for matrix \textbf{A}$^0$ since it contains purely auxiliary parameters.\\

\noindent \textbf{USD Spot Pairs:} Table \ref{tab:vLogMEM1_USDspot} reports the estimates for matrices \textbf{A} and \textbf{B} of the vLogMEM(1,1) fitted to diurnally-adjusted five-minute realised volatility on the three USD spot pairs.\footnote{Since we include only one lag in our LogMEM, we omit the indices of the two matrices \textbf{A}$_1$ and \textbf{B}$_1$.}$^{,}$\footnote{For reasons of better readability, we report non-significance rather than significance of the parameter estimates.} It also includes the model's log-likelihood values and BIC. We can see that the estimates on Coinbase change only marginally compared to the univariate results in Table \ref{tab:uniMEM_results} -- the short-term persistence $\alpha$ decreases from 0.37 to about 0.34, while the long-term persistence level $\beta$ reduces only from 0.59 to 0.58 -- indicating that the volatility on Coinbase is rather independent of Bitstamp and Kraken.\footnote{By short-term persistence, we mean the diagonal entries of the matrix \textbf{A}, i.e. the volatility flows of an instrument to itself. Based on how much this parameter changes compared to the univariate analysis, we can draw conclusions about the dependencies among instruments.} By contrast, the levels of persistence on Bitstamp and Kraken change substantially once lagged volatility of Coinbase is included, which indicates a certain degree of dependence on the latter. On Bitstamp, short-term (long-term) persistence reduces from 0.38 (0.60) to 0.28 (0.55), while the estimates on Kraken drop from 0.38 and 0.60 to about 0.21 and 0.50.

The non-diagonal entries of matrix \textbf{A} which capture the lagged dependence among exchanges reveal strong volatility flows from Coinbase to Bitstamp and Kraken but not vice versa. While the estimates for spillovers from Coinbase to Bitstamp and Kraken are highly significant at 0.12 and 0.20, the reversed flows are much less (0.05 and 0.004, respectively) and in the case of Kraken, they are not even significant. Putting these numbers in simple economic terms, a 29\%-shock in realised volatility on Coinbase (i.e. $\varepsilon_t = 1.2$) -- which is the conditional standard deviation of the innovations on Coinbase assuming only strictly positive values -- leads to a relative increase of about 3.1\% (5.2\%) in expected volatility on Bitstamp (Kraken).\footnote{Note that we use the log of the conditional mean in the LogMEM, therefore we first have to apply the exponential function to obtain the conditional mean itself. Moreover, for the calculation of the standard deviation, we assume there were only strictly positive volatility values, i.e. the innovations follow a simple log-normal distribution. Otherwise, we would have to take \textbf{A}$_0$ into account as well, but since these are purely auxiliary parameters necessary to model the volatility at high frequency and the impact of zero volatility values is not the object of this study, we assume only strictly positive values.}$^, $\footnote{Also note that this amount is not the complete increase. Since the contemporaneous correlation among innovations, which our model does not capture, is still quite high even at the five-minute frequency, the actual increase will be a little more.} On the other hand, a one-standard deviation shock on Bitstamp -- again assuming only non-zero volatility values -- increases the expected volatility on Coinbase only by 1\%. 
Among Bitstamp and Kraken, we detect similar unidirectional volatility flows. Spillovers from Bitstamp to Kraken are highly significant at 0.13, while the reversed flows are quite weak and only about 0.02. That is, a one-standard deviation volatility shock on Bitstamp leads to a relative increase of 3.2\% in expected volatility on Kraken. By contrast, a one-standard deviation shock on Kraken raises the expected volatility on Bitstamp only marginally by less than 1\%.

Following these results, we conclude that within the fiat-margined spot market, Coinbase is the main source of volatility. It exhibits major flows to both Bitstamp and Kraken and receives only little volatility from Bitstamp. Among the two smaller exchanges, Bitstamp and Kraken, the former is the larger emitter of volatility, albeit its flows to Kraken are significantly smaller than those from Coinbase. Finally, Kraken transmits only very marginal flows to Bitstamp and is the main receiver of volatility.


\begin{table}[tb]
 	\centering
 	\small
 	\caption{Multivariate LogMEM(1,1) -- USD Spot}
    \begin{tabular}{lrlll}
          &       & \multicolumn{1}{c}{\textbf{Coinbase}} & \multicolumn{1}{c}{\textbf{Bitstamp}} & \multicolumn{1}{c}{\textbf{Kraken}} \\
          \midrule
    \multirow{3}{*}{\textbf{A}}     & \multicolumn{1}{l}{Coinbase} & \hphantom{$-$}{\color{c2}0.3383} & \hphantom{$-$}0.0546 & \hphantom{$-$}0.0044$^{ns}$ \\
          & \multicolumn{1}{l}{Bitstamp} & \hphantom{$-$}0.1184 & \hphantom{$-$}{\color{c2}0.2846} & \hphantom{$-$}0.0248 \\
          & \multicolumn{1}{l}{Kraken} & \hphantom{$-$}0.2006 & \hphantom{$-$}0.1285 & \hphantom{$-$}{\color{c2}0.2073} \\
          &       &       &       &  \\
    \multirow{3}{*}{\textbf{B}}     & \multicolumn{1}{l}{Coinbase} & \hphantom{$-$}0.5760 &       &  \\
          & \multicolumn{1}{l}{Bitstamp} &       & \hphantom{$-$}0.5522 &  \\
          & \multicolumn{1}{l}{Kraken} &       &       & \hphantom{$-$}0.4975 \\
    \midrule
    LL & & \multicolumn{1}{c}{$-$653} & \multicolumn{1}{c}{$-$20,962} & \multicolumn{1}{c}{$-$25,418} \\
    BIC & & \multicolumn{1}{c}{1,408} & \multicolumn{1}{c}{ \hphantom{$-$}42,025} & \multicolumn{1}{c}{\hphantom{$-$}50,937} \\
    \midrule
    \end{tabular}%
	{\raggedright  \justify\footnotesize \textit{Note:} The table reports parameter estimates, log-likelihood value and BIC for the  vLogMEM(1,1), fitted to five-minute realised volatility on Coinbase (CB), Bitstamp (BS) and Kraken (KK) over the period from 1 January to 31 March 2021. The superscript $^{ns}$ indicates that the estimate is not significant at the 1\%-level. For better readibility, the parameters capturing short-term persistence are highlighted in red.\par}
 	\label{tab:vLogMEM1_USDspot}
\end{table}%

\ \\
\noindent \textbf{USDT Spot Pairs:} Table \ref{tab:vLogMEM1_USDTspot} reports parameter estimates, log-likelihood values and BIC for the two tether spot pairs on Binance and Huobi. Similarly to the fiat-margined analysis above, we see a significant reduction in short-term persistence on Binance -- it drops from 0.39 in the univariate analysis to less than 0.26. Huobi on the other hand experiences only a slight reduction in short-term persistence from 0.39 to about 0.36, which suggests that realised volatility on Huobi is rather independendent of Binance. This is also confirmed by the non-diagonal entries of matrix \textbf{A}. With a highly significant estimate of 0.14, volatility flows from Huobi to Binance are very strong, whereas the reversed flows from Binance to Huobi are not even significant. Therefore, we conclude that within the tether-margined spot market, volatility mainly emerges on Huobi from where it then spills over to Binance. 

\begin{table}[tb]
  \centering
  \small
  \caption{Multivariate LogMEM(1,1) -- USDT Spot}
    \begin{tabular}{llll}
          &       & \multicolumn{1}{c}{\textbf{Binance$^S$}} & \multicolumn{1}{c}{\textbf{Huobi}} \\
    \midrule
    \multirow{2}[0]{*}{\textbf{A}} & \multicolumn{1}{l}{Binance$^S$} & \hphantom{$-$}{\color{c2}0.2591} & \hphantom{$-$}0.1377 \\
          & \multicolumn{1}{l}{Huobi} & \hphantom{$-$}0.0282$^{ns}$ & \hphantom{$-$}{\color{c2}0.3586} \\
          &       &       &  \\
    \multirow{2}[0]{*}{\textbf{B}} & \multicolumn{1}{l}{Binance$^S$} & \hphantom{$-$}0.5679 &  \\
          & \multicolumn{1}{l}{Huobi} &       & \hphantom{$-$}0.5780 \\
    \midrule
    LL & & \multicolumn{1}{c}{\hphantom{$-$1,}556} & \multicolumn{1}{c}{\hphantom{$-$}1,316} \\
    BIC & & \multicolumn{1}{c}{$-$1,030} & \multicolumn{1}{c}{$-$2,551} \\
    \midrule
    \end{tabular}
	{\raggedright \footnotesize \justify \textit{Note:} The table reports parameter estimates, log-likelihood and BIC for the  vLogMEM(1,1), fitted to five-minute realised volatility on Binance Spot (BI$^S$) and Huobi (HU) over the period from 1 January to 31 March 2021. The superscript $^{ns}$ indicates that the estimate is not significant at the 1\%-level. For better readibility, the parameters capturing short-term persistence are highlighted in red. \par}
 	\label{tab:vLogMEM1_USDTspot}
\end{table}

\ \\
\noindent \textbf{Perpetual Swaps:} Table \ref{tab:vLogMEM1_perp} reports parameter estimates, log-likelihood value and BIC for the three perpetual swaps. We see that the dynamics of the Binance tether-margined perpetual are rather independent of the other two contracts, with its parameter estimates showing almost no changes compared to the univariate analysis. By contrast, the two fiat-margined perpetuals on Binance and Bybit experience significant parameter changes -- especially their short-term persistence reduces substantially from 0.29 to 0.14 for Bybit and from 0.38 to 0.25 for Binance. On Bybit, the long-term persistence also drops significantly to 0.61. These results suggest that realised volatility on the two USD perpetuals on Binance and Bybit depends strongly on the tether-margined Binance contract.

The non-diagonal entries of matrix \textbf{A} indicate strong unidirectional flows from the Binance USDT perpetual to the two USD-margined products. With estimates of 0.18 and 0.15, the Binance tether-contract transmits a large amount of volatility to both the Bybit and Binance fiat-perpetual -- on Bybit, its influence even exceeds the short-term persistence. Our estimates imply that a one-standard deviation volatility shock on the USDT-product leads to a relative increase of 4.3\% (3.4\%) in expected volatility on the Bybit (Binance) USD-contract. By contrast, the Binance tether-margined perpetual does not receive any significant volatility flows, as can be seen from the two non-significant entries in the first row of \textbf{A}. Among the two USD perpetuals, Binance seems to be more important in terms of volatility transmission. Spillovers from Binance to Bybit are estimated as 0.09, while the reversed flows are not significant at the 1\%-level. 

Based on these results, we conclude that the main source of volatility within the perpetuals market is the tether-margined contract on Binance. From there, large amounts of volatility are transmitted to the two largest fiat-margined perpetuals on Bybit and Binance. In addition, there are smaller volatility flows from the Binance USD-contract to Bybit. The fact that Bybit only receives, but does not transmit volatility might again be related to its market structure. Both fees and contract specifications make the product highly attractive for retail traders, generating rather uninformed trading activity to which traders on Binance do not react. Moreover, it might increase the time that traders on Bybit need to pick up signals from other exchanges, leading to a lagged dependence on realised volatility of Binance.

\begin{table}[tb]
  \centering
  \small
  \caption{Multivariate LogMEM(1,1) -- Perpetuals}
    \begin{tabular}{lllll}
          &       & \multicolumn{1}{c}{\textbf{Binance$^T$}} & \multicolumn{1}{c}{\textbf{Bybit}} & \multicolumn{1}{c}{\textbf{Binance$^{\$}$}} \\
          \midrule
    \multirow{3}[0]{*}{\textbf{A}} &   \multicolumn{1}{l}{Binance$^T$}    & \hphantom{$-$}{\color{c2}0.3939} & \hphantom{$-$}0.0126$^{ns}$ & $-$0.0089$^{ns}$ \\
          &   \multicolumn{1}{l}{Bybit}    & \hphantom{$-$}0.1831 & \hphantom{$-$}{\color{c2}0.1402} & \hphantom{$-$}0.0867 \\
          &     \multicolumn{1}{l}{Binance$^{\$}$}  & \hphantom{$-$}0.1462 & \hphantom{$-$}0.0083$^{ns}$ & \hphantom{$-$}{\color{c2}0.2471} \\
          &       &       &       &  \\
    \multirow{3}[0]{*}{\textbf{B}} &   \multicolumn{1}{l}{Binance$^T$}    & \hphantom{$-$}0.5620 &       &  \\
          &  \multicolumn{1}{l}{Bybit}     &      & \hphantom{$-$}0.6113 &  \\
          &  \multicolumn{1}{l}{Binance$^{\$}$}     &       &       & \hphantom{$-$}0.5753 \\
     \midrule
    LL & & \multicolumn{1}{c}{\hphantom{$-$}1,450} & \multicolumn{1}{c}{\hphantom{1}$-$8,114} & \multicolumn{1}{c}{$-$1,549}\\
    BIC & & \multicolumn{1}{c}{$-$2,799} & \multicolumn{1}{c}{\hphantom{$-$}16,331} & \multicolumn{1}{c}{\hphantom{$-$}3,199}\\
    \midrule
    \end{tabular}
	{\raggedright \footnotesize \justify \textit{Note:} The table reports parameter estimates, log-likelihood and BIC for the vLogMEM(1,1), fitted to five-minute realised volatility on the Binance USDT perpetual (BI$^T$), Bybit perpetual (BY) and Binance USD perpetual (BI$^{\$}$) over the period from 1 January to 31 March 2021. The superscript $^{ns}$ indicates that the estimate is not significant at the 1\%-level. For better readibility, the parameters capturing short-term persistence are highlighted in red. \par}
  	\label{tab:vLogMEM1_perp}
\end{table}

\ \\
\noindent \textbf{Main Instruments:} Table \ref{tab:vLogMEM1_major} reports the estimates for matrices \textbf{A} and \textbf{B} of the six-dimensional vLogMEM(1,1) fitted to five-minute realised volatility on Coinbase, Huobi, Bybit and the three Binance instruments. It also includes the model's log-likelihood values and BIC.\footnote{To detect potential time-variation, we have also estimated the model for each of the three months separately. The overall results however do not change significantly, indicating that the realised volatility dynamics are quite stable over time.} As before, the degree of short-term persistence changes significantly compared to the univariate dynamics. The most extreme change occurs on the Binance USDT spot pair -- the estimate reduces from 0.39 to 0.07 -- while the Binance tether-perpetual is the only instrument where the inclusion of lagged interdependence leads to an increase in short-term persistence. As in the previous multivariate analyses, the remaining parameters (asymmetric response component, long-term persistence, standard deviation of log residuals) do not change significantly. Only on Bybit, the level of long-term persistence drops slightly from 0.68 in the univariate analysis to about 0.62. 

Overall, the Binance USDT perpetual has the second lowest level of long-term persistence -- only on the Binance USDT spot pair, it is marginally smaller -- and by far the highest degree of short-term persistence. Therefore, the result from the univariate analysis that traders on this contract are more sensitive to prevailing market conditions than on the remaining five instruments still holds, once we allow for lagged interdependence.

For better illustration, Figure \ref{fig:flow_plot} shows the magnitude of volatility flows (i.e. the non-diagonal entries of matrix \textbf{A}) as a circular plot. Note that non-significant entries are set to zero. From this figure it becomes clear that the Binance USDT perpetual is the main source of volatility. It exhibits very strong spillovers to all other instruments, while it only receives quite weak volatility flows from Coinbase and Binance spot. The two remaining perpetuals (Bybit and Binance USD) are far less important and besides some volatility flows between the two products, there is only a minor spillover from Bybit to Huobi. Interestingly, both Coinbase and Binance spot transmit volatility to all other instruments. However, as can be seen in Table \ref{tab:vLogMEM1_major}, all of these flows are negative. An explanation for this might be that traders on these spot exchanges need longer to react to (temporarily) increased volatility on the perpetual swaps. Once the volatility has risen on the spot exchanges, the transient increase on the perpetual contracts is already reversed, leading to a negative estimate of the volatility transmitted from spot exchanges. Finally, both Bybit and Huobi seem to be receivers rather than transmitters of volatility. Besides a weak spillover between the two instruments and minor flows from Huobi to Binance spot, we detect no significant influence on the remaining four instruments.

\begin{table}[tb]
  \centering
  \small
  \caption{Multivariate LogMEM(1,1) -- Main Instruments}
    \begin{tabular}{lllllllll}
          &       & \multicolumn{1}{c}{\textbf{Coinbase}} & \multicolumn{1}{c}{\textbf{Binance$^S$}} & \multicolumn{1}{c}{\textbf{Huobi}} & \multicolumn{1}{c}{\textbf{Binance$^T$}} & \multicolumn{1}{c}{\textbf{Bybit}} & \multicolumn{1}{c}{\textbf{Binance$^\$$}} & \multicolumn{1}{|c}{To}\\
          \midrule
    \multirow{6}[0]{*}{\textbf{A}} & \multicolumn{1}{l}{Coinbase} & \hphantom{$-$}{\color{c2}0.2108} & $-$0.0895 & \hphantom{$-$}0.0352$^{ns}$ & \hphantom{$-$}0.2548 & \hphantom{$-$}0.0088$^{ns}$ & $-$0.0173$^{ns}$ & \multicolumn{1}{|l}{0.3762}\\
          & \multicolumn{1}{l}{Binance$^S$} & $-$0.0164$^{ns}$ & \hphantom{$-$}{\color{c2}0.0742} & \hphantom{$-$}0.0626 & \hphantom{$-$}0.2439 & \hphantom{$-$}0.0105$^{ns}$ & \hphantom{$-$}0.0223$^{ns}$ & \multicolumn{1}{|l}{0.3806} \\
          & \multicolumn{1}{l}{Huobi} & $-$0.0282 & $-$0.0851 & \hphantom{$-$}{\color{c2}0.2720} & \hphantom{$-$}0.2177 & \hphantom{$-$}0.0198 & $-$0.0108$^{ns}$ & \multicolumn{1}{|l}{0.3962} \\
          & \multicolumn{1}{l}{Binance$^T$} & $-$0.0307 & $-$0.0995 & \hphantom{$-$}0.0247$^{ns}$ & \hphantom{$-$}{\color{c2}0.4702} & \hphantom{$-$}0.0149$^{ns}$ & \hphantom{$-$}0.0132$^{ns}$ & \multicolumn{1}{|l}{0.3400}\\
          & \multicolumn{1}{l}{Bybit} & $-$0.0761 & $-$0.1373 & \hphantom{$-$}0.0441$^{ns}$ & \hphantom{$-$}0.3085 & \hphantom{$-$}{\color{c2}0.1450} & \hphantom{$-$}0.1200 & \multicolumn{1}{|l}{0.3600} \\
          & \multicolumn{1}{l}{Binance$^\$$} & $-$0.0594 & $-$0.1039 & \hphantom{$-$}0.0110$^{ns}$ & \hphantom{$-$}0.2606 & \hphantom{$-$}0.0134$^{ns}$ & \hphantom{$-$}{\color{c2}0.2742} & \multicolumn{1}{|l}{0.3716} \\
          \hline
          & From & \hphantom{$-$}0.0163 & $-$0.4411	& \hphantom{$-$}0.3346 & \hphantom{$-$}1.7557 & \hphantom{$-$}0.1648 & \hphantom{$-$}0.3943 & \multicolumn{1}{|l}{ } \\
          &       &       &       &       &       &       &  & \\
    \multirow{6}[0]{*}{\textbf{B}} & \multicolumn{1}{l}{Coinbase} & \hphantom{$-$}0.5749 &       &       &       &       &  & \\
          & \multicolumn{1}{l}{Binance$^S$} &       & \hphantom{$-$}0.5651 &       &       &       &  & \\
          & \multicolumn{1}{l}{Huobi} &       &       & \hphantom{$-$}0.5780 &       &       &  & \\
          & \multicolumn{1}{l}{Binance$^T$} &       &       &       & \hphantom{$-$}0.5665 &       &  & \\
          & \multicolumn{1}{l}{Bybit} &       &       &       &       & \hphantom{$-$}0.6167 &  & \\
          & \multicolumn{1}{l}{Binance$^\$$} &       &       &       &       &       & \hphantom{$-$}0.5805 & \\
    \midrule
    LL & & \multicolumn{1}{c}{$-$562} & \multicolumn{1}{c}{\hphantom{$-$1,}690} & \multicolumn{1}{c}{\hphantom{$-$}1,425} & \multicolumn{1}{c}{\hphantom{$-$}1,482} & \multicolumn{1}{c}{$-$8,072} & \multicolumn{1}{c}{$-$1,504} & \\
    BIC & & \multicolumn{1}{c}{1,287} & \multicolumn{1}{c}{$-$1,218} & \multicolumn{1}{c}{$-$2,687} & \multicolumn{1}{c}{$-$2,801} & \multicolumn{1}{c}{16,307} & \multicolumn{1}{c}{\hphantom{$-$}3,170} & \\
    \midrule
    \end{tabular}
	{\raggedright \footnotesize \justify \textit{Note:} The table reports parameter estimates, log-likelihood and BIC for the vLogMEM(1,1), fitted to five-minute realised volatility on Coinbase (CB), Binance Spot (BI$^S$), Huobi (HU), Binance USDT perpetual (BI$^T$), Bybit (BY) and Binance USD perpetual (BI$^{\$}$) over the period from 1 January to 31 March 2021. The superscript $^{ns}$ indicates that the estimate is not significant at the 1\%-level. For better readibility, the parameters capturing short-term persistence are highlighted in red. \par}
  	\label{tab:vLogMEM1_major}
\end{table}

\subsection{Intra-day Variation of Spillovers}\label{subsec:spillover_timeofday}
Similar to the FX market, crypto exchanges operate globally and are open 24/7. But trading activity varies significantly over the course of the day, as shown by Figures \ref{fig:intra-day_volume} and \ref{fig:intra-day_volatility}. Therefore it is of particular importance to examine the volatility dynamics for short-term intra-day variation.\footnote{Intra-day time-variation of the long-term persistence is rather unlikely and therefore, we focus on variation in the short-term parameters.} We do so by decomposing the trading day into three different time periods of equal length, namely the Asian trading hours from midnight to 08:00 UTC, the European time period from 08:00 to 16:00 UTC and finally the U.S.  trading hours from 16:00 to midnight UTC. For each of these three time periods, we introduce in our multivariate LogMEM(1,1) an interaction term of the log realised volatility with a dummy variable that is one during the respective time period and zero otherwise. Consequently, the conditional mean specification in \eqref{eq:vLogMEM1} changes to
\begin{equation*}
	\begin{split}
	\log \boldsymbol{\mu}_t = \textbf{w} &+ \left( \textbf{A} + \textbf{A}_{\text{AS}} \mathds{1}_{\text{AS}} + \textbf{A}_{\text{EU}} \mathds{1}_{\text{EU}} + \textbf{A}_{\text{US}} \mathds{1}_{\text{US}} \right)  \left( \log \boldsymbol{x}_{t-1} \odot \mathds{1}_{\{ \boldsymbol{x}_{t-j}>0 \}} \right) \\
	&+ \textbf{A}^0 \mathds{1}_{\{ \boldsymbol{x}_{t-1}=0 \}} + \boldsymbol{\Gamma} \log \boldsymbol{x}_{t-1}^- + \textbf{B} \log \boldsymbol{\mu}_{t-1}
	\end{split}
\end{equation*}
where $\mathds{1}_{AS}$, $\mathds{1}_{EU}$ and $\mathds{1}_{US}$ denote the dummy variables for Asian, European and U.S.  trading hours, respectively. All other variables are the same as in \eqref{eq:vLogMEM1}.

\begin{table}[!htb]
	\centering
	\small
	\caption{Multivariate LogMEM(1,1) -- Intra-day}
	\begin{tabular}{llllllll}
		&       & \multicolumn{1}{c}{\textbf{Coinbase}} & \multicolumn{1}{c}{\textbf{Binance$^S$}} & \multicolumn{1}{c}{\textbf{Huobi}} & \multicolumn{1}{c}{\textbf{Binance$^T$}} & \multicolumn{1}{c}{\textbf{Bybit}} & \multicolumn{1}{c}{\textbf{Binance$^\$$}} \\
		\midrule
		\multirow{6}[0]{*}{\textbf{A}} & Coinbase & \hphantom{$-$}{\color{c2}0.1906}$^{***}$ & \multicolumn{1}{c}{--} & \hphantom{$-$}0.0715$^{*}$ & \multicolumn{1}{c}{--} & \multicolumn{1}{c}{--} & \hphantom{$-$}0.1082$^{***}$ \\
		& Binance$^S$ & \multicolumn{1}{c}{--} & \hphantom{$-$}{\color{c2}0.1226}$^{**}$ & \hphantom{$-$}0.1119$^{***}$ & \multicolumn{1}{c}{--} & \multicolumn{1}{c}{--} & \hphantom{$-$}0.1046$^{***}$ \\
		& Huobi & $-$0.0505$^{**}$ & \multicolumn{1}{c}{--} & \hphantom{$-$}{\color{c2}0.3514}$^{***}$ & \multicolumn{1}{c}{--} & \multicolumn{1}{c}{--} & \hphantom{$-$}0.0781$^{**}$ \\
		& Binance$^T$ & $-$0.0456$^{**}$ & \multicolumn{1}{c}{--} & \hphantom{$-$}0.0646$^{*}$ & \hphantom{$-$}{\color{c2}0.2669}$^{***}$ & \multicolumn{1}{c}{--} & \hphantom{$-$}0.1095$^{***}$ \\
		& Bybit & $-$0.0934$^{***}$ & \multicolumn{1}{c}{--} & \multicolumn{1}{c}{--} & \hphantom{$-$}0.1671$^{**}$ & \hphantom{$-$}{\color{c2}0.0435}$^{**}$ & \hphantom{$-$}0.3117$^{***}$ \\
		& Binance$^\$$ & $-$0.0779$^{***}$ & \multicolumn{1}{c}{--} & \multicolumn{1}{c}{--} & \hphantom{$-$}0.1277$^{**}$ & \multicolumn{1}{c}{--} & \hphantom{$-$}{\color{c2}0.3431}$^{***}$ \\
		&       &       &       &       &       &       &  \\
		\multirow{6}[0]{*}{\textbf{A}$_{\text{AS}}$} & Coinbase & \multicolumn{1}{c}{--} & \multicolumn{1}{c}{--} & \multicolumn{1}{c}{--} & \hphantom{$-$}0.1546$^{**}$ & \hphantom{$-$}0.0826$^{***}$ & $-$0.1902$^{***}$ \\
		& Binance$^S$ & \multicolumn{1}{c}{--} & \multicolumn{1}{c}{--} & \multicolumn{1}{c}{--} & \hphantom{$-$}0.1172$^{**}$ & \hphantom{$-$}0.0577$^{***}$ & $-$0.1087$^{**}$ \\
		& Huobi & \multicolumn{1}{c}{--} & \multicolumn{1}{c}{--} & \multicolumn{1}{c}{--} & \hphantom{$-$}0.1325$^{**}$ & \hphantom{$-$}0.0658$^{***}$ & $-$0.1343$^{***}$ \\
		& Binance$^T$ & \multicolumn{1}{c}{--} & \multicolumn{1}{c}{--} & \multicolumn{1}{c}{--} &\multicolumn{1}{c}{--} & \hphantom{$-$}0.0720$^{***}$ & $-$0.1420$^{***}$ \\
		& Bybit & \multicolumn{1}{c}{--} & \multicolumn{1}{c}{--} & \multicolumn{1}{c}{--} & \multicolumn{1}{c}{--} & \hphantom{$-$}{\color{c2}0.2177}$^{***}$ & $-$0.2702$^{***}$ \\
		& Binance$^\$$ & \multicolumn{1}{c}{--}& \multicolumn{1}{c}{--} & \multicolumn{1}{c}{--} & \multicolumn{1}{c}{--} & \hphantom{$-$}0.0729$^{***}$ & $-${\color{c2}0.1108}$^{*}$ \\
		&       &       &       &       &       &       &  \\
		\multirow{6}[0]{*}{\textbf{A}$_{\text{EU}}$} & Coinbase & \multicolumn{1}{c}{--} & $-$0.0978$^{**}$ & \multicolumn{1}{c}{--} & \hphantom{$-$}0.1813$^{***}$ & \hphantom{$-$}0.0546$^{**}$ & $-$0.1466$^{***}$ \\
		& Binance$^S$ & \multicolumn{1}{c}{--} & $-$0.0984$^{*}$ & \multicolumn{1}{c}{--} & \hphantom{$-$}0.1402$^{**}$ & \hphantom{$-$}0.0462$^{**}$ & \multicolumn{1}{c}{--} \\
		& Huobi & \multicolumn{1}{c}{--} & \multicolumn{1}{c}{--} & {\color{c2}$-$0.0689}$^{*}$ & \hphantom{$-$}0.1701$^{***}$ & \hphantom{$-$}0.0486$^{**}$ & $-$0.1025$^{*}$ \\
		& Binance$^T$ &\multicolumn{1}{c}{--} & $-$0.1198$^{***}$ & \multicolumn{1}{c}{--} & \hphantom{$-$}{\color{c2}0.1569}$^{***}$ & \hphantom{$-$}0.0469$^{**}$ & $-$0.0855$^{*}$ \\
		& Bybit & \multicolumn{1}{c}{--} & $-$0.1356$^{***}$ & \multicolumn{1}{c}{--} & \hphantom{$-$}0.1447$^{**}$ & \hphantom{$-$}{\color{c2}0.1885}$^{***}$ & $-$0.2818$^{***}$ \\
		& Binance$^\$$ & \multicolumn{1}{c}{--} & $-$0.1043$^{**}$ & \multicolumn{1}{c}{--} & \multicolumn{1}{c}{--} & \multicolumn{1}{c}{--} & \multicolumn{1}{c}{--} \\
		&       &       &       &       &       &       &  \\
		\multirow{6}[0]{*}{\textbf{A}$_{\text{US}}$} & Coinbase & \multicolumn{1}{c}{--} & \multicolumn{1}{c}{--} & \multicolumn{1}{c}{--} & \hphantom{$-$}0.2221$^{***}$ & \hphantom{$-$}0.0781$^{***}$ & $-$0.1656$^{***}$ \\
		& Binance$^S$ & \multicolumn{1}{c}{--} & \multicolumn{1}{c}{--} & $-$0.0848$^{**}$ & \hphantom{$-$}0.1654$^{***}$ & \hphantom{$-$}0.0610$^{***}$ & $-$0.1073$^{**}$ \\
		& Huobi & \multicolumn{1}{c}{--} & $-$0.0935$^{*}$ & \multicolumn{1}{c}{--} & \hphantom{$-$}0.2020$^{***}$ & \hphantom{$-$}0.0572$^{***}$ & $-$0.0956$^{*}$ \\
		& Binance$^T$ & \multicolumn{1}{c}{--} & \multicolumn{1}{c}{--} & \multicolumn{1}{c}{--} & \hphantom{$-$}{\color{c2}0.1902}$^{***}$ & \hphantom{$-$}0.0611$^{***}$ & $-$0.1148$^{**}$ \\
		& Bybit & \multicolumn{1}{c}{--} & $-$0.1215$^{**}$ & \multicolumn{1}{c}{--}& \hphantom{$-$}0.1903$^{***}$ & \hphantom{$-$}{\color{c2}0.1978}$^{***}$ & $-$0.2793$^{***}$ \\
		& Binance$^\$$ &\multicolumn{1}{c}{--} & \multicolumn{1}{c}{--} & \multicolumn{1}{c}{--} & \hphantom{$-$}0.1257$^{**}$ & \hphantom{$-$}0.0627$^{***}$ & \multicolumn{1}{c}{--} \\
		\midrule
		LL & & \multicolumn{1}{c}{$-$502} & \multicolumn{1}{c}{\hphantom{$-$1,}760} & \multicolumn{1}{c}{\hphantom{$-$}1,489} & \multicolumn{1}{c}{\hphantom{$-$}1,565} & \multicolumn{1}{c}{$-$8,012} & \multicolumn{1}{c}{$-$1,450} \\
		BIC & & \multicolumn{1}{c}{1,350} & \multicolumn{1}{c}{$-$1,175} & \multicolumn{1}{c}{$-$2,633} & \multicolumn{1}{c}{$-$2,785} & \multicolumn{1}{c}{16,370} & \multicolumn{1}{c}{\hphantom{$-$}3,246} \\
		\midrule
	\end{tabular}
	{\raggedright \footnotesize \justify \textit{Note:} The table reports parameter estimates, log-likelihood and BIC for the vLogMEM(1,1), fitted to five-minute realised volatility on Coinbase (CB), Binance Spot (BI$^S$), Huobi (HU), Binance USDT perpetual (BI$^T$), Bybit (BY) and Binance USD perpetual (BI$^{\$}$) over the period from 1 January to 31 March 2021. The asterisks $^{***}$,$^{**}$, $^{*}$ indicate significance at the 1\%, 5\%, and 10\% level, respectively, based on robust standard errors. For better readibility, entries that are not significant at the 10\%-level are already removed and the parameters capturing short-term persistence are highlighted in red. \par}
	\label{tab:vLogMEM1_major_intra-day}
\end{table}

Table \ref{tab:vLogMEM1_major_intra-day} reports the resulting estimates for the short-term matrices \textbf{A}, \textbf{A}$_\text{AS}$, \textbf{A}$_\text{EU}$ and \textbf{A}$_\text{US}$, together with log-likelihood values and BIC. 
For reasons of clarity and comprehensibility, entries that are not significant at the 10\%-level are already removed. We do not include the estimates for the remaining parameters since most of them do not change significantly -- only the intercepts on Coinbase, Huobi and the three Binance instruments reduce slightly to about $-$0.05. This shows that volatility flows from the spot exchanges (Coinbase, Binance$^S$, Huobi) are far more constant over the course of the day than those from the three perpetual swaps. In particular, flows transmitted by Coinbase do not exhibit any intra-day variation and the estimates from matrix \textbf{A} are quite close to those in the previous analysis. Similarly, we do not find significant intra-day variation on Huobi, where only two entries -- the short-term persistence during European trading and the volatility flows to Binance spot during the U.S.  trading period -- are negative and (weakly) significant. The Binance USDT spot pair on the other hand only transmits volatility during European trading hours and, to a lesser extent, during U.S.  trading times. The parameter estimates in the second column of \textbf{A}$_\text{EU}$ are very similar to the short-term estimates in Table \ref{tab:vLogMEM1_major} and therefore, we conclude that the previously documented volatility flows from the Binance USDT spot pair are mainly caused by European trading activity.

All three perpetual swaps exhibit a high degree of intra-day variation, with only eight out of the 54 associated parameters of the interaction matrices \textbf{A}$_\text{AS}$, \textbf{A}$_\text{EU}$ and \textbf{A}$_\text{US}$ being not significant. As in the previous analysis, the Binance USDT perpetual exhibits the strongest volatility flows to all other instruments. However, these flows intensify throughout the day and reach their maximum during U.S.  trading hours, when the spillovers to all other instruments are (highly) significant and (far) above 0.16. The Bybit and Binance USD perpetuals both exhibit significant volatility spillovers over the whole day, but their intra-day patterns differ from the tether-margined contract. While the total volatility flows transmitted by Bybit are largest (lowest) in Asian (European) trading, the Binance USD perpetual emits most volatility during European trading hours and only rather little in Asian trading. 
We conclude that even after accounting for intra-day variation in volatility spillovers, the Binance tether-margined contract is still the instrument that generates and transmits most of the volatility. However, its leadership role is less pronounced depending on the time of day and the Bybit perpetual in particular gains relative importance during Asian trading.

Overall, the volatility flows among the six crypto instruments seem to strengthen over the course of the day. While the total spillover amount -- measured as the sum of absolute flows, excluding the short-term persistence parameter -- is 1.78 in Asian trading, it increases by 30\% to about 2.30 during European and U.S.  trading hours. Also, when considering only the short-term persistence (i.e. the diagonal entries of the matrices), we find a similar intra-day pattern. In Asian and European trading, the total short-term persistence of the six instruments is about 1.43 and 1.50, respectively, while it amounts to more than 1.70 during U.S.  trading hours. That is, a volatility shock that occurs during the U.S.  trading period raises the traders' expectations on the realised volatility within the next five-minute interval significantly more than a shock during Asian or European trading times. These two findings suggest that (i) during U.S.  trading hours, traders pay more attention to prevailing market conditions when updating their expectations and (ii) the crypto market exhibits a higher interconnectedness when traditional Western stock markets are open, which is consistent with Figure \ref{fig:intra-day_volume}.

\subsection{Robustness Checks}\label{subsec:spillover_robustness}
To check the robustness of our volatility transmission results, we repeat the above analysis using two alternative estimators of realised volatility. In particular, we use the Bipower Variation (BPV) of \citet{BarndorffNielsen2004, BarndorffNielsen2006} calculated as
\begin{equation}\label{eq:BPV}
	BPV = \frac{\pi}{2} \sum_{j=2}^{n} \lvert r_{t+j-1} \rvert \lvert r_{t+j} \rvert
\end{equation} 
and the Median Realised Volatility (MedRV) by \citet{Andersen2012} which is given by
\begin{equation}\label{eq:MedRV}
	MedRV = \frac{\pi}{6-4\sqrt{3}+\pi} \sum_{j=2}^{n-1} \text{med}(\lvert r_{t+j-1} \rvert, \lvert r_{t+j} \rvert, \lvert r_{t+j+1} \rvert)^2.
\end{equation}
Both of these measures are robust towards (infrequent) jumps in the return series -- for the BPV, a jump may be offset by an adjacent small return, while the application of the median in the MedRV reduces the effect of an isolated, very large return to zero.\footnote{Please note that equations \eqref{eq:BPV} and \eqref{eq:MedRV} calculate the realised variance and we need to take the square root and multiply it by $\sqrt{365 \times 288}$ to obtain the realised volatility.}  In addition, due to the use of the median, the MedRV estimator is also particularly robust towards zero returns. Therefore, these two measures are well suited to check the robustness of our results towards the choice of realised volatility estimator.\footnote{In addition to these two alternative measures, we also repeated the above analyses using simple squared log returns as realised volatility estimator. The results however are qualitatively similar to the pre-averaged volatility measure. Therefore, we do not report them here but they are available upon request.}

We proceed in exactly the same way as before, i.e. we first remove the intra-day pattern -- which looks very similar to the pattern observed for the pre-averaged volatility -- and winsorize the top 0.05\% of the diurnally-adjusted realised volatility values. Then, we estimate both univariate and multivariate versions of the LogMEM – first for each group of instruments separately and then for the most important instruments jointly. 
For reasons of brevity, we report only selected results here, but all results are available from the authors upon request.\footnote{It is worth noting that we could not estimate the LogMEM for Kraken. In particular, the persistence parameters $\alpha$ and $\beta$ converged quickly to estimates very similar to the analyses using the pre-averaged realised volatiliy. However, the three remaining parameters ($\omega$, $\alpha_0$ and $s$) diverged to (minus) infinity. This behaviour is probably caused by an unusual extremity in the realised volatility. Due to low trading activity on Kraken, the time series exhibits quite a lot of zero values. At the same time however, we obtain extreme non-zero volatility values since we do not pre-average the one-second returns. The kurtosis of the diurnally-adjusted and winsorized realised volatility on Kraken is 39 (BPV) and 40 (MedRV) -- compared to ``only'' 9 for the pre-averaged measure. Due to this extremity, the LogMEM probably reaches its limits and we do exclude Kraken from our robustness analysis.} For most products, both MedRV and BPV are slightly more (less) short-term (long-term) persistent which is probably caused by the microstructure-reducing local smoothing applied in the pre-averaged volatility estimator. However, for all multivariate analyses, the results are consistent with those obtained using the pre-averaged realised volatility. Most importantly, as can be seen in Tables \ref{tab:vLogMEM1_major_bpvRV} and \ref{tab:vLogMEM1_major_medRV}, the multivariate analysis including the main instruments still leads to the conclusion that the Binance USDT-margined perpetual contract is the main source of volatility. It transmits significant, strong flows to all other instruments and receives only very little volatility from Coinbase and Bybit.

Interestingly, in contrast to the analysis using pre-averaged realised volatility, the volatility flows from Coinbase are either not statistically significant or slightly positive. 
This difference might be related to microstructure noise. The pre-averaged volatility estimator smooths the one-second returns locally to reduce the impact of noise to a certain extent -- the MedRV and BPV do not explicitly do this. Therefore, the now positive spillovers from Coinbase, the best-known crypto exchange and thus the main trading venue for less informed retail traders, might be due to microstructure noise that is transmitted to the other instruments.

\begin{table}[tb]
	\centering
	\small
	\caption{Multivariate LogMEM(1,1) -- Main Instruments; Bipower Variation}
	\begin{tabular}{lllllllll}
		&       & \multicolumn{1}{c}{\textbf{Coinbase}} & \multicolumn{1}{c}{\textbf{Binance$^S$}} & \multicolumn{1}{c}{\textbf{Huobi}} & \multicolumn{1}{c}{\textbf{Binance$^T$}} & \multicolumn{1}{c}{\textbf{Bybit}} & \multicolumn{1}{c}{\textbf{Binance$^\$$}} & \multicolumn{1}{|c}{To}\\
		\midrule
		\multirow{6}[0]{*}{\textbf{A}} & \multicolumn{1}{l}{Coinbase} & \hphantom{$-$}{\color{c2}0.3263} & $-$0.0063$^{ns}$ & $-$0.0465 & \hphantom{$-$}0.0931 & \hphantom{$-$}0.0401 & \hphantom{$-$}0.0329 & \multicolumn{1}{|l}{0.4459}\\
		& \multicolumn{1}{l}{Binance$^S$} & \hphantom{$-$}0.0207 & \hphantom{$-$}{\color{c2}0.2669} & $-$0.0009$^{ns}$ & \hphantom{$-$}0.0991 & \hphantom{$-$}0.0353 & \hphantom{$-$}0.0264 & \multicolumn{1}{|l}{0.4484} \\
		& \multicolumn{1}{l}{Huobi} & $-$0.0029$^{ns}$ & $-$0.0223$^{ns}$ & \hphantom{$-$}{\color{c2}0.3439} & \hphantom{$-$}0.0425 & \hphantom{$-$}0.0458 & \hphantom{$-$}0.0076$^{ns}$ & \multicolumn{1}{|l}{0.4321} \\
		& \multicolumn{1}{l}{Binance$^T$} & \hphantom{$-$}0.0227 & \hphantom{$-$}0.0133$^{ns}$ & \hphantom{$-$}0.0078$^{ns}$ & \hphantom{$-$}{\color{c2}0.3532} & \hphantom{$-$}0.0343 & \hphantom{$-$}0.0197$^{ns}$ & \multicolumn{1}{|l}{0.4102}\\
		& \multicolumn{1}{l}{Bybit} & \hphantom{$-$}0.0187$^{ns}$ & $-$0.0069$^{ns}$ & \hphantom{$-$}0.0327$^{ns}$ & \hphantom{$-$}0.0812 & \hphantom{$-$}{\color{c2}0.2073} & \hphantom{$-$}0.0631 & \multicolumn{1}{|l}{0.3515} \\
		& \multicolumn{1}{l}{Binance$^\$$} & \hphantom{$-$}0.0294 & \hphantom{$-$}0.0187$^{ns}$ & $-$0.0083$^{ns}$ & \hphantom{$-$}0.1213 & \hphantom{$-$}0.0656 & \hphantom{$-$}{\color{c2}0.2230} & \multicolumn{1}{|l}{0.4394} \\
		\hline
		& From & \hphantom{$-$}0.3991 & \hphantom{$-$}0.2669 & \hphantom{$-$}0.2973 & \hphantom{$-$}0.7903 & \hphantom{$-$}0.4283 & \hphantom{$-$}0.3455 & \multicolumn{1}{|l}{ } \\			
		&       &       &       &       &       &       &  & \\
		\multirow{6}[0]{*}{\textbf{B}} & \multicolumn{1}{l}{Coinbase} & \hphantom{$-$}0.5154 &       &       &       &       &  & \\
		& \multicolumn{1}{l}{Binance$^S$} &       & \hphantom{$-$}0.4779 &       &       &       &  & \\
		& \multicolumn{1}{l}{Huobi} &       &       & \hphantom{$-$}0.5282 &       &       &  & \\
		& \multicolumn{1}{l}{Binance$^T$} &       &       &       & \hphantom{$-$}0.4699 &       &  & \\
		& \multicolumn{1}{l}{Bybit} &       &       &       &       & \hphantom{$-$}0.6116 &  & \\
		& \multicolumn{1}{l}{Binance$^\$$} &       &       &       &       &       & \hphantom{$-$}0.5308 & \\
		\midrule
		LL & & \multicolumn{1}{c}{$-$2,841} & \multicolumn{1}{c}{\hphantom{0}$-$99} & \multicolumn{1}{c}{\hphantom{0}$-$992} & \multicolumn{1}{c}{$-$281} & \multicolumn{1}{c}{\hphantom{0}$-$9,791} & \multicolumn{1}{c}{\hphantom{0}$-$6,674} & \\
		BIC & & \multicolumn{1}{c}{\hphantom{$-$}5,844} & \multicolumn{1}{c}{\hphantom{$-$}360} & \multicolumn{1}{c}{\hphantom{$-$}2,146} & \multicolumn{1}{c}{$-$400} & \multicolumn{1}{c}{\hphantom{$-$}19,744} & \multicolumn{1}{c}{\hphantom{$-$}13,511} & \\
		\midrule
	\end{tabular}
	{\raggedright \footnotesize \justify \textit{Note:} The table reports parameter estimates, log-likelihood and BIC for the vLogMEM(1,1), fitted to five-minute Bipower Variation on Coinbase (CB), Binance Spot (BI$^S$), Huobi (HU), Binance USDT perpetual (BI$^T$), Bybit (BY) and Binance USD perpetual (BI$^{\$}$) over the period from 1 January to 31 March 2021. The superscript $^{ns}$ indicates that the estimate is not significant at the 1\%-level. For better readibility, the parameters capturing short-term persistence are highlighted in red. \par}
	\label{tab:vLogMEM1_major_bpvRV}
\end{table}

\begin{table}[tb]
	\centering
	\small
	\caption{Multivariate LogMEM(1,1) -- Main Instruments; Median Realised Volatility}
	\begin{tabular}{lllllllll}
		&       & \multicolumn{1}{c}{\textbf{Coinbase}} & \multicolumn{1}{c}{\textbf{Binance$^S$}} & \multicolumn{1}{c}{\textbf{Huobi}} & \multicolumn{1}{c}{\textbf{Binance$^T$}} & \multicolumn{1}{c}{\textbf{Bybit}} & \multicolumn{1}{c}{\textbf{Binance$^\$$}} & \multicolumn{1}{|c}{To}\\
		\midrule
		\multirow{6}[0]{*}{\textbf{A}} & \multicolumn{1}{l}{Coinbase} & \hphantom{$-$}{\color{c2}0.3208} & \hphantom{$-$}0.0043$^{ns}$ & $-$0.0497 & \hphantom{$-$}0.0797 & \hphantom{$-$}0.0425 & \hphantom{$-$}0.0359 & \multicolumn{1}{|l}{0.4292}\\
		& \multicolumn{1}{l}{Binance$^S$} & \hphantom{$-$}0.0195 & \hphantom{$-$}{\color{c2}0.2744} & $-$0.0051$^{ns}$ & \hphantom{$-$}0.1045 & \hphantom{$-$}0.0341 & \hphantom{$-$}0.0217$^{ns}$ & \multicolumn{1}{|l}{0.4325} \\
		& \multicolumn{1}{l}{Huobi} & $-$0.0059$^{ns}$ & $-$0.0198$^{ns}$ & \hphantom{$-$}{\color{c2}0.3505} & \hphantom{$-$}0.0524 & \hphantom{$-$}0.0416 & \hphantom{$-$}0.0023$^{ns}$ & \multicolumn{1}{|l}{0.4446} \\
		& \multicolumn{1}{l}{Binance$^T$} & \hphantom{$-$}0.0190 & \hphantom{$-$}0.0187$^{ns}$ & \hphantom{$-$}0.0127$^{ns}$ & \hphantom{$-$}{\color{c2}0.3474} & \hphantom{$-$}0.0365 & \hphantom{$-$}0.0157$^{ns}$ & \multicolumn{1}{|l}{0.4029}\\
		& \multicolumn{1}{l}{Bybit} & \hphantom{$-$}0.0266 & $-$0.0167$^{ns}$ & \hphantom{$-$}0.0188$^{ns}$ & \hphantom{$-$}0.1161 & \hphantom{$-$}{\color{c2}0.2098} & \hphantom{$-$}0.0586 & \multicolumn{1}{|l}{0.4111} \\
		& \multicolumn{1}{l}{Binance$^\$$} & \hphantom{$-$}0.0277 & \hphantom{$-$}0.0174$^{ns}$ & $-$0.0015$^{ns}$ & \hphantom{$-$}0.1278 & \hphantom{$-$}0.0612 & \hphantom{$-$}{\color{c2}0.2327} & \multicolumn{1}{|l}{0.4494} \\
		\hline
		& From & \hphantom{$-$}0.4135 & \hphantom{$-$}0.2744 & \hphantom{$-$}0.3009 & \hphantom{$-$}0.8280 & \hphantom{$-$}0.4258 & \hphantom{$-$}0.3272 & \multicolumn{1}{|l}{ } \\			
		&       &       &       &       &       &       &  & \\
		\multirow{6}[0]{*}{\textbf{B}} & \multicolumn{1}{l}{Coinbase} & \hphantom{$-$}0.5153 &       &       &       &       &  & \\
		& \multicolumn{1}{l}{Binance$^S$} &       & \hphantom{$-$}0.4759 &       &       &       &  & \\
		& \multicolumn{1}{l}{Huobi} &       &       & \hphantom{$-$}0.5217 &       &       &  & \\
		& \multicolumn{1}{l}{Binance$^T$} &       &       &       & \hphantom{$-$}0.4713 &       &  & \\
		& \multicolumn{1}{l}{Bybit} &       &       &       &       & \hphantom{$-$}0.6073 &  & \\
		& \multicolumn{1}{l}{Binance$^\$$} &       &       &       &       &       & \hphantom{$-$}0.5284 & \\
		\midrule
		LL & & \multicolumn{1}{c}{$-$2,347} & \multicolumn{1}{c}{\hphantom{$-$}261} & \multicolumn{1}{c}{$-$214} & \multicolumn{1}{c}{\hphantom{$-$0,}798} & \multicolumn{1}{c}{$-$10,250} & \multicolumn{1}{c}{\hphantom{0}$-$6,173} & \\
		BIC & & \multicolumn{1}{c}{\hphantom{$-$}4,857} & \multicolumn{1}{c}{$-$358} & \multicolumn{1}{c}{\hphantom{$-$}590} & \multicolumn{1}{c}{$-$1,433} & \multicolumn{1}{c}{\hphantom{$-$}20,663} & \multicolumn{1}{c}{\hphantom{$-$}12,508} & \\
		\midrule
	\end{tabular}
	{\raggedright \footnotesize \justify \textit{Note:} The table reports parameter estimates, log-likelihood and BIC for the vLogMEM(1,1), fitted to five-minute Median Realised Volatility on Coinbase (CB), Binance Spot (BI$^S$), Huobi (HU), Binance USDT perpetual (BI$^T$), Bybit (BY) and Binance USD perpetual (BI$^{\$}$) over the period from 1 January to 31 March 2021. The superscript $^{ns}$ indicates that the estimate is not significant at the 1\%-level. For better readibility, the parameters capturing short-term persistence are highlighted in red. \par}
	\label{tab:vLogMEM1_major_medRV}
\end{table}

\section{Volume Results}\label{sec:volume}

\subsection{Volume Flows}
We proceed exactly in the same way as before, i.e. we first adjust trading volume for its intra-day pattern and winsorize the top 0.05\% to reduce the influence of extreme outliers. Then we estimate the LogMEM -- first the univariate version to all instruments separately and then the multivariate model to each instrument group, and fianlly to our selection of six main instruments. We also estimate the model including dummy interaction terms to detect intra-day patterns, as in Section \ref{subsec:spillover_timeofday}. To economize on space, we only report  selected results here -- deltailed results are available from the authors upon request. 

Table \ref{tab:vLogMEM1_major:volume} reports the matrices \textbf{A} and \textbf{B} resulting from the vLogMEM(1,1) fitted to five-minute trading volume on Coinbase, Huobi, Bybit and the three Binance instruments. The results are quite different from the volatility flows reported in Table \ref{tab:vLogMEM1_major}. Trading volume has much more short-term persistence than realised volatility -- the diagonal entries of \textbf{A} and \textbf{B} are much higher and lower, respectively -- indicating that volume is much more sensitive to the current state of the market. This finding is is in line with the results of \citet{Nguyen2020} on US Treasury notes. 

More importantly, however, Table \ref{tab:vLogMEM1_major:volume} shows that volume flows between the instruments on different exchanges have little relationship with volatility spillovers. In particular, now it is the spot pairs, especially on Coinbase and Binance,  that transmit the most volume to the perpetuals. Coinbase transmits over three times as much volume as the Binance tether-margined perpetual, which is the main source of realised volatility. Interestingly, among the three perpetual contracts, the Bybit contract transmits most volume even though it is the least relevant in terms of volatility generation and transmission. By contrast, the Binance USD perpetual does not transmit volume to any of the other instruments. 

Clearly, our previous results on volatility transmission are not simply related to volume transmission. In fact, taken together we find that it is not only volume which matters for volatility transmission, but also the quality of information in this volume. We conclude that the two spot pairs on Coinbase and Binance, as well as the Bybit perpetual contract, are  the instruments with the least informed traders. This is because Coinbase and Binance are the most widely known spot exchanges and the properties of the Bybit contract make it attractive for smaller, retail investors. While these exchanges transmit the most trading volume, in terms of volatility transmission they are much less relevant than the Binance tether-margined perpetual.

\begin{table}[tb]
	\centering
	\small
	\caption{Multivariate LogMEM(1,1) for Trading Volume -- Main Instruments}
	\begin{tabular}{lllllllll}
		&       & \multicolumn{1}{c}{\textbf{Coinbase}} & \multicolumn{1}{c}{\textbf{Binance$^S$}} & \multicolumn{1}{c}{\textbf{Huobi}} & \multicolumn{1}{c}{\textbf{Binance$^T$}} & \multicolumn{1}{c}{\textbf{Bybit}} & \multicolumn{1}{c}{\textbf{Binance$^\$$}}&  \multicolumn{1}{|c}{To}\\
		\midrule
		\multirow{6}[0]{*}{\textbf{A}} & \multicolumn{1}{l}{Coinbase}  & \hphantom{$-$}{\color{c2}0.4094} & \hphantom{$-$}0.0664 & \hphantom{$-$}0.0195 & $-$0.0576 & \hphantom{$-$}0.0175 & $-$0.0058$^{ns}$ & \multicolumn{1}{|l}{0.4551}\\
		& \multicolumn{1}{l}{Binance$^S$} & \hphantom{$-$}0.0768 & \hphantom{$-$}{\color{c2}0.4114} & \hphantom{$-$}0.0175 & $-$0.0529 & \hphantom{$-$}0.0208 & \hphantom{$-$}0.0110$^{ns}$ & \multicolumn{1}{|l}{0.4735}\\
		& \multicolumn{1}{l}{Huobi} & \hphantom{$-$}0.0880 & \hphantom{$-$}0.0960 & \hphantom{$-$}{\color{c2}0.3825} & $-$0.0572 & \hphantom{$-$}0.0076$^{ns}$	& $-$0.0135$^{ns}$ & \multicolumn{1}{|l}{0.5093}\\
		& \multicolumn{1}{l}{Binance$^T$} & \hphantom{$-$}0.0619 & \hphantom{$-$}0.0546 & $-$0.0039$^{ns}$ & \hphantom{$-$}{\color{c2}0.3538} & \hphantom{$-$}0.0454 & $-$0.0019$^{ns}$ & \multicolumn{1}{|l}{0.5157}\\
		& \multicolumn{1}{l}{Bybit} & \hphantom{$-$}0.0561 & \hphantom{$-$}0.0537 & $-$0.0272 & \hphantom{$-$}0.0377 & \hphantom{$-$}{\color{c2}0.3642} & $-$0.0065$^{ns}$ & \multicolumn{1}{|l}{0.4846}\\
		& \multicolumn{1}{l}{Binance$^\$$} & \hphantom{$-$}0.0569 & \hphantom{$-$}0.0660 & $-$0.0057$^{ns}$ & $-$0.0119$^{ns}$ & \hphantom{$-$}0.0468 & \hphantom{$-$}{\color{c2}0.3176} & \multicolumn{1}{|l}{0.4874}\\
		\hline
		& From & \hphantom{$-$}0.7491 & \hphantom{$-$}0.7481	& \hphantom{$-$}0.3923 & \hphantom{$-$}0.2238 & \hphantom{$-$}0.4947 & \hphantom{$-$}0.3176 & \multicolumn{1}{|l}{ } \\
		&       &       &       &       &       &       &  & \\
		\multirow{6}[0]{*}{\textbf{B}} & \multicolumn{1}{l}{Coinbase} & \hphantom{$-$}0.4954 &       &       &       &       &  & \\
		& \multicolumn{1}{l}{Binance$^S$} &       & \hphantom{$-$}0.3915 &       &       &       &  & \\
		& \multicolumn{1}{l}{Huobi} &       &       & \hphantom{$-$}0.3789 &       &       &  & \\
		& \multicolumn{1}{l}{Binance$^T$} &       &       &       & \hphantom{$-$}0.3725 &       &  & \\
		& \multicolumn{1}{l}{Bybit} &       &       &       &       & \hphantom{$-$}0.4467 &  & \\
		& \multicolumn{1}{l}{Binance$^\$$} &       &       &       &       &       & \hphantom{$-$}0.4509 & \\
		\midrule
		LL & & \multicolumn{1}{l}{\hphantom{1}$-$9,525} & \multicolumn{1}{l}{\hphantom{1}$-$8,500} & \multicolumn{1}{l}{$-$12,245} & \multicolumn{1}{l}{$-$10,817} & \multicolumn{1}{l}{$-$13,307} & \multicolumn{1}{l}{$-$12,172} & \\
		BIC & & \multicolumn{1}{l}{\hphantom{$-$}19,193} & \multicolumn{1}{l}{\hphantom{$-$}17,143} & \multicolumn{1}{l}{\hphantom{$-$}24,632} & \multicolumn{1}{l}{\hphantom{$-$}21,777} & \multicolumn{1}{l}{\hphantom{$-$}26,755} & \multicolumn{1}{l}{\hphantom{$-$}24,486} & \\
		\midrule
	\end{tabular}
	{ \raggedright \footnotesize \justify \textit{Note:} The table reports parameter estimates, log-likelihood and BIC for the vLogMEM(1,1), fitted to five-minute trading volume (in USD) on Coinbase (CB), Binance Spot (BI$^S$), Huobi (HU), Binance USDT perpetual (BI$^T$), Bybit (BY) and Binance USD perpetual (BI$^{\$}$) over the period from 1 January to 31 March 2021. The superscript $^{ns}$ indicates that the estimate is not significant at the 1\%-level. For better readibility, the parameters capturing short-term persistence are highlighted in red. \par}
	\label{tab:vLogMEM1_major:volume}
\end{table}

To further corroborate that our results on realised volatility transmission are different from volume spillovers, Table \ref{tab:vLogMEM1_major_intraday} reports results of estimating the vLogMEM(1,1) with dummy interaction terms to capture intra-day variation. Compared with realised volatility, the five-minute trading volume is rather stable over the course of the day, as most entries of the matrix \textbf{A} are highly significant while the dummy matrices \textbf{A}$_\text{AS}$, \textbf{A}$_\text{EU}$ and \textbf{A}$_\text{US}$ have many non-significant elements. In fact, we observe almost no intra-day volume variation on Huobi, Bybit or the Binance instruments. Only the volume flows from Coinbase change significantly: they are lowest during Asian trading hours and substantially increase during European and US trading. Overall however, the two spot pairs on Coinbase and Binance are still the main transmitters of trading volume, even after accounting for intra-day variation. 
In Table \ref{tab:vLogMEM1_major_intraday}, it is also worth noting that all instruments except for the Binance inverse perpetual exhibit an elevated short-term persistence (i.e. the diagonal entries of the matrices \textbf{A}, \textbf{A}$_\text{AS}$, \textbf{A}$_\text{EU}$ and \textbf{A}$_\text{US}$) during European and/or US trading hours which once more indicates that  traders pay more attention to prevailing market conditions during these times.

\begin{table}[!tb]
	\centering
	\small
	\caption{Multivariate LogMEM(1,1) for Trading Volume -- Intra-day}
	\begin{tabular}{llllllll}
		&       & \multicolumn{1}{c}{\textbf{Coinbase}} & \multicolumn{1}{c}{\textbf{Binance$^S$}} & \multicolumn{1}{c}{\textbf{Huobi}} & \multicolumn{1}{c}{\textbf{Binance$^T$}} & \multicolumn{1}{c}{\textbf{Bybit}} & \multicolumn{1}{c}{\textbf{Binance$^\$$}} \\
		\midrule
		
		\multirow{6}[0]{*}{\textbf{A}} & Coinbase & \hphantom{$-$}{\color{c2}0.3873}$^{***}$ & \hphantom{$-$}0.0548$^{***}$ & \multicolumn{1}{c}{--} & $-$0.0503$^{***}$	& \hphantom{$-$}0.0188$^{**}$ & \multicolumn{1}{c}{--} \\
		& Binance$^S$ & \hphantom{$-$}0.0447$^{***}$ & \hphantom{$-$}{\color{c2}0.3891}$^{***}$ & \multicolumn{1}{c}{--} & $-$0.0717$^{***}$ & \hphantom{$-$}0.0166$^{**}$ & \hphantom{$-$}0.0416$^{***}$ \\
		& Huobi & \hphantom{$-$}0.0435$^{***}$ & \hphantom{$-$}0.0911$^{***}$ & \hphantom{$-$}{\color{c2}0.3562}$^{***}$ & $-$0.0523$^{***}$ & \multicolumn{1}{c}{--} & \multicolumn{1}{c}{--} \\
		& Binance$^T$ & \hphantom{$-$}0.0329$^{***}$ & \hphantom{$-$}0.0486$^{***}$ & $-$0.0186$^{*}$ & \hphantom{$-$}{\color{c2}0.3138}$^{***}$ & \hphantom{$-$}0.0481$^{***}$ & \multicolumn{1}{c}{--} \\
		& Bybit & \hphantom{$-$}0.0383$^{***}$ & \hphantom{$-$}0.0557$^{***}$ & $-$0.0537$^{***}$ & \hphantom{$-$}0.0407$^{*}$ & \hphantom{$-$}{\color{c2}0.3337}$^{***}$ & \multicolumn{1}{c}{--} \\
		& Binance$^\$$ & \hphantom{$-$}0.0270$^{***}$ & \hphantom{$-$}0.0911$^{***}$ & $-$0.0183$^{*}$ & $-$0.0358$^{*}$ & \hphantom{$-$}0.0464$^{***}$ & \hphantom{$-$}{\color{c2}0.3193}$^{***}$ \\
		&       &       &       &       &       &       &  \\
		
		\multirow{6}[0]{*}{\textbf{A}$_{\text{AS}}$} & Coinbase & \multicolumn{1}{c}{--} & \multicolumn{1}{c}{--} & \hphantom{$-$}0.0191$^{**}$ & \multicolumn{1}{c}{--} & \multicolumn{1}{c}{--} & \hphantom{$-$}0.0281$^{*}$ \\
		& Binance$^S$ & \hphantom{$-$}0.0258$^{***}$ & \multicolumn{1}{c}{--} & \multicolumn{1}{c}{--} & \multicolumn{1}{c}{--} & \multicolumn{1}{c}{--} & \multicolumn{1}{c}{--} \\
		& Huobi &  \hphantom{$-$}0.0395$^{***}$ & \multicolumn{1}{c}{--} & \multicolumn{1}{c}{--} & $-$0.0402$^{*}$ & \multicolumn{1}{c}{--} & \multicolumn{1}{c}{--} \\
		& Binance$^T$ & \multicolumn{1}{c}{--} & \multicolumn{1}{c}{--} & \multicolumn{1}{c}{--} & \multicolumn{1}{c}{--} & \multicolumn{1}{c}{--} & \multicolumn{1}{c}{--} \\
		& Bybit &   \multicolumn{1}{c}{--} &  \multicolumn{1}{c}{--} & \hphantom{$-$}0.0377$^{***}$ &  \multicolumn{1}{c}{--} & \hphantom{$-$}{\color{c2}0.0250}$^{**}$ &  \multicolumn{1}{c}{--}\\
		& Binance$^\$$ & \hphantom{$-$}0.0214$^{**}$ & \multicolumn{1}{c}{--} & \hphantom{$-$}0.0168$^{*}$ & \multicolumn{1}{c}{--} & \multicolumn{1}{c}{--} & \multicolumn{1}{c}{--} \\
		&       &       &       &       &       &       &  \\
		
		\multirow{6}[0]{*}{\textbf{A}$_{\text{EU}}$} & Coinbase & \hphantom{$-$}{\color{c2}0.0271}$^{***}$ & \multicolumn{1}{c}{--} & \multicolumn{1}{c}{--} & \multicolumn{1}{c}{--} & \multicolumn{1}{c}{--} & \multicolumn{1}{c}{--} \\
		& Binance$^S$ & \hphantom{$-$}0.0356$^{***}$ & \hphantom{$-$}{\color{c2}0.0250}$^{*}$ & \multicolumn{1}{c}{--} & \hphantom{$-$}0.0329$^{*}$ & \multicolumn{1}{c}{--} & $-$0.0539$^{***}$ \\
		& Huobi & \hphantom{$-$}0.0493$^{***}$ & \multicolumn{1}{c}{--} & \multicolumn{1}{c}{--}  & \multicolumn{1}{c}{--}  & \multicolumn{1}{c}{--}  & \multicolumn{1}{c}{--}  \\
		& Binance$^T$ & \hphantom{$-$}0.0391$^{***}$ & \multicolumn{1}{c}{--} & \multicolumn{1}{c}{--} & \hphantom{$-$}{\color{c2}0.0670}$^{***}$ & $-$0.0139$^{*}$ & $-$0.0507$^{***}$ \\
		& Bybit & \hphantom{$-$}0.0311$^{***}$ & \multicolumn{1}{c}{--} & \multicolumn{1}{c}{--} & \multicolumn{1}{c}{--} & \hphantom{$-$}{\color{c2}0.0272}$^{***}$ & $-$0.0398$^{**}$ \\
		& Binance$^\$$ & \hphantom{$-$}0.0430$^{***}$ & \multicolumn{1}{c}{--} & \multicolumn{1}{c}{--} & \hphantom{$-$}0.0414$^{**}$ & \multicolumn{1}{c}{--} & \multicolumn{1}{c}{--} \\
		&       &       &       &       &       &       &  \\
		
		\multirow{6}[0]{*}{\textbf{A}$_{\text{US}}$} & Coinbase & \hphantom{$-$}{\color{c2}0.0309}$^{***}$ & \multicolumn{1}{c}{--} & \multicolumn{1}{c}{--} & \multicolumn{1}{c}{--} & \multicolumn{1}{c}{--} & \multicolumn{1}{c}{--} \\
		& Binance$^S$ & \hphantom{$-$}0.0432$^{***}$ & \multicolumn{1}{c}{--} & \multicolumn{1}{c}{--} & \hphantom{$-$}0.0340$^{*}$ & \multicolumn{1}{c}{--} & $-$0.0348$^{**}$ \\
		& Huobi & \hphantom{$-$}0.0591$^{***}$ & $-$0.0392$^{**}$ & \hphantom{$-$}{\color{c2}0.0274}$^{***}$ & \multicolumn{1}{c}{--} & \multicolumn{1}{c}{--} & \multicolumn{1}{c}{--} \\
		& Binance$^T$ & \hphantom{$-$}0.0385$^{***}$ & \multicolumn{1}{c}{--} & \multicolumn{1}{c}{--} & \hphantom{$-$}{\color{c2}0.0599}$^{***}$ & \multicolumn{1}{c}{--} & \multicolumn{1}{c}{--} \\
		& Bybit & \hphantom{$-$}0.0290$^{***}$ & \multicolumn{1}{c}{--} & \hphantom{$-$}0.0214$^{**}$ & \multicolumn{1}{c}{--} & \hphantom{$-$}{\color{c2}0.0370}$^{***}$ & \multicolumn{1}{c}{--} \\
		& Binance$^\$$ & \hphantom{$-$}0.0349$^{***}$ & $-$0.0499$^{***}$ & \multicolumn{1}{c}{--} & \multicolumn{1}{c}{--} & \multicolumn{1}{c}{--} & \multicolumn{1}{c}{--} \\
		\midrule
		LL & & \multicolumn{1}{l}{\hphantom{1}$-$9,492} & \multicolumn{1}{l}{\hphantom{1}$-$8,434} & \multicolumn{1}{l}{$-$12,174} & \multicolumn{1}{l}{$-$10,757} & \multicolumn{1}{l}{$-$13,244} & \multicolumn{1}{l}{$-$12,128} \\
		BIC & & \multicolumn{1}{l}{\hphantom{$-$}19,310} & \multicolumn{1}{l}{\hphantom{$-$}17,194} & \multicolumn{1}{l}{\hphantom{$-$}24,674} & \multicolumn{1}{l}{\hphantom{$-$}21,838} & \multicolumn{1}{l}{\hphantom{$-$}26,814} & \multicolumn{1}{l}{\hphantom{$-$}24,581} \\
		\midrule
	\end{tabular}
	{\raggedright \footnotesize \justify \textit{Note:} The table reports parameter estimates, log-likelihood and BIC for the vLogMEM(1,1), fitted to five-minute trading volume (in USD) on Coinbase (CB), Binance Spot (BI$^S$), Huobi (HU), Binance USDT perpetual (BI$^T$), Bybit (BY) and Binance USD perpetual (BI$^{\$}$) over the period from 1 January to 31 March 2021. The asterisks $^{***}$,$^{**}$, $^{*}$ indicate significance at the 1\%, 5\%, and 10\% level, respectively, based on robust standard errors. For better readibility, entries that are not significant at the 10\%-level are already removed and the parameters capturing short-term persistence are highlighted in red. \par}
	\label{tab:vLogMEM1_major_intraday}
\end{table}

\subsection{Volume-Volatility Relationship}\label{subsec:volume_volatility_relationship}

To study the interaction between trading volume and realised volatility, we estimate six bivariate LogMEM for Coinbase, Huobi, Bybit and the three Binance instruments. Both trading volume and realised volatility are diurnally-adjusted to remove the deterministic intra-day patterns documented in Figures \ref{fig:intra-day_volume} and \ref{fig:intra-day_volatility}. In addition, we winsorize the top 0.05\% of the two time series. In contrast with the previous analyses, we include two lags in the short-term persistence (i.e. we fit a LogMEM(2,1)) thus increasing the model’s flexibility and ability to capture interactions between the two variables.

\begin{table}[tb]
  \centering
  \small
  \caption{Bivariate LogMEM(2,1) -- Trading Volume \& Realised Volatility}
    \begin{tabular}{c|l|ll|ll|ll}
          &       & \multicolumn{2}{c|}{Coinbase} & \multicolumn{2}{c|}{Binance Spot} & \multicolumn{2}{c}{Huobi} \\
          &       & \multicolumn{1}{c}{TV} & \multicolumn{1}{c|}{RV} & \multicolumn{1}{c}{TV} & \multicolumn{1}{c|}{RV} & \multicolumn{1}{c}{TV} & \multicolumn{1}{c}{RV} \\
    \midrule
    \multirow{2}[2]{*}{\textbf{A}$_1$} & TV    & \hphantom{$-$}0.4953 & $-$0.0548 & \hphantom{$-$}0.5560 & $-$0.1123 & \hphantom{$-$}0.4935 & \multicolumn{1}{l}{$-$0.0796} \\
          & RV    & \hphantom{$-$}0.0966 & \hphantom{$-$}0.3394 & \hphantom{$-$}0.1364 & \hphantom{$-$}0.3039 & \hphantom{$-$}0.0790 & \multicolumn{1}{l}{\hphantom{$-$}0.3471} \\
    \midrule
    \multirow{2}[1]{*}{\textbf{A}$_2$} & TV    & $-$0.3048 & \hphantom{$-$}0.0547 & $-$0.3754 & \hphantom{$-$}0.1123 & $-$0.3171 & \multicolumn{1}{l}{\hphantom{$-$}0.0995} \\
          & RV    & $-$0.0946 & $-$0.1891 & $-$0.1360 & $-$0.1523 & $-$0.0778 & \multicolumn{1}{l}{$-$0.1923} \\
    \multicolumn{1}{r}{} & \multicolumn{1}{r}{} &       & \multicolumn{1}{r}{} &       & \multicolumn{1}{r}{} &       &  \\
    \multicolumn{1}{r}{} & \multicolumn{1}{r}{} &       & \multicolumn{1}{r}{} &       & \multicolumn{1}{r}{} &       &  \\
          &       & \multicolumn{2}{c|}{Binance USDT} & \multicolumn{2}{c|}{Bybit} & \multicolumn{2}{c}{Binance USD} \\
          &       & \multicolumn{1}{c}{TV} & \multicolumn{1}{c|}{RV} & \multicolumn{1}{c}{TV} & \multicolumn{1}{c|}{RV} & \multicolumn{1}{c}{TV} & \multicolumn{1}{c}{RV} \\
    \midrule
    \multirow{2}[2]{*}{\textbf{A}$_1$} & TV    & \hphantom{$-$}0.5526 & $-$0.1100 & \hphantom{$-$}0.5157 & $-$0.1116 & \hphantom{$-$}0.4606 & $-$0.0045$^{ns}$\\
          & RV    & \hphantom{$-$}0.1076 & \hphantom{$-$}0.3023 & \hphantom{$-$}0.1384 & \hphantom{$-$}0.2208 & \hphantom{$-$}0.0990 & \hphantom{$-$}0.2988 \\
    \midrule
    \multirow{2}[1]{*}{\textbf{A}$_2$} & TV    & $-$0.4141 & \hphantom{$-$}0.1072 & $-$0.3555 & \hphantom{$-$}0.1047 & $-$0.3101 & \hphantom{$-$}0.0015$^{ns}$ \\
          & RV    & $-$0.1112 & $-$0.1216$^{ns}$ & $-$0.1396 & $-$0.0985 & $-$0.1035 & $-$0.1267 \\
          \midrule
    \end{tabular}
	{\raggedright \footnotesize \justify \textit{Note:} The table reports the parameter estimates for the matrices \textbf{A}$_1$ and \textbf{A}$_2$ from the bivariate LogMEM(2,1), fitted to five-minute trading volume (TV) and realised volatility (RV) on Coinbase, Binance Spot, Huobi, Binance USDT perpetual, Bybit and Binance USD perpetual over the period from 1 January to 31 March 2021. The superscript $^{ns}$ indicates that the estimate is not significant at the 10\%-level. \par}
  	\label{tab:volume_volatility}
\end{table}

The resulting estimates of the six bivariate models are reported in Table \ref{tab:volume_volatility}. As before, we only report the matrices \textbf{A}$_1$ and \textbf{A}$_2$ since the remaining parameters are less relevant for our analysis. We see that the relationship between trading volume and volatility is similar for all of the six instruments.
In the first lag, the impact of volume on volatility is positive, while that of volatility on volume is negative. This implies that an increase in volatility is preceded by an increase in trading volume in the previous five-minute interval. However, in the second lag, the relationship is reversed, leading to a negative effect of volume on volatility and a positive influence of volatility on volume. We suppose that this inversion is caused by some mean-reverting behaviour -- indicated by the negative diagonal entries of \textbf{A}$_2$ -- and a longer response time (i.e. a time delay) of volatility. This is also supported by the diagonal entries of the matrix \textbf{A}$_1$ which imply a higher short-term persistence of volume than volatility. Overall, we conclude that there is a significant bi-directional causality relationship between trading volume and realised volatility. Except, for Binance USD perpetual, we find that trading volume causes realised volatility, but the effect of volatility on volume is not significant in either lag.

\section{Conclusions}\label{sec:concl} 
We analyse high-frequency realised volatility dynamics and spillovers in the bitcoin market during the most recent bull period from 1 January to 31 March 2021. Our analysis focuses on two cryptocurrency pairs, namely trading bitcoin against the U.S.  dollar (the main fiat-crypto pair) and trading bitcoin against tether (the main crypto-crypto pair). Based on second-by-second transaction data covering the major spot and perpetual exchanges, we estimate both univariate and multivariate versions of the Logarithmic Multiplicative Error Model -- first for each group of instruments separately and then for the most important instruments jointly. Finally, we examine the inter-exchange spillovers for intra-day variation by dividing the day into three different time zones of equal length (Asian, European, U.S.  trading hours). 
The tether-margined perpetual contract on Binance receives the least volatility flows during Asian, European and U.S.  trading time zones. However, it is clearly the main emitter of volatility.  
During European and U.S.  trading, the total spillover amount is about 30\% higher than during Asian trading hours, when  
traders on this contract are the most reactive to prevailing market conditions. 
and exchanges exhibits a higher inter-connectedness. We also show that volatility transmission patterns differ markedly from volume flows, where Coinbase is the main source of transmission. 

The results and discussions in this paper aim to illuminate tether's ability for contagion and the role of Binance in volatility transmission within the bitcoin market. Binance is not just a trading venue, it acts as a broker and custodian of tether, and a large fraction of the recent tether issuance can be linked to Binance's demand.\footnote{For instance,  on 31 May Paolo Ardoini, CTO of Tether and Bitfinex, \href{https://www.reddit.com/user/StablecoinsFraud/comments/o0u5cg/tether_chain_swap_fraud_non_existent_burns/}{tweeted} that Binance ordered \$3bn more tether to be minted, and one of Binance cold wallets held about 17.1 bn tokens, almost 30\% of the total tether supply, at the beginning of July 2021. See the unofficial \href{https://wallet.tether.to/richlist}{Tether Rich List}.}  Thus, our concern about the  dominance of Binance is exacerbated by an equal concern about tether's ever-increasing market capitalization and its growing usage. 
Binance could also be implicated in the long-standing concerns about tether's collateralization. The stable coin’s issuer, Tether Limited, originally claimed to hold one U.S.  dollar for each token of tether. But in May 2021, after recurring accusations of creating un-backed tokens and a prolonged investigation by the New York Attorney General, Tether reported that only 2.9\% of all tokens are actually backed by cash reserves and about 50\% is in commercial paper, a form of unsecured debt that is normally only issued by firms with high-quality debt ratings. 
The simultaneous growth of Binance and tether even begs the question whether Binance itself is the issuer of a large fraction of tether’s \$30 billion commercial paper. 
While the speculation surrounding the issuer(s) will continue as long as Tether remains unwilling to undergo a reliable third-party audit,\footnote{See \href{https://www.ft.com/content/529eb4e6-796a-4e81-8064-5967bbe3b4d9}{Financial Times} on 14 May 2021 and the \href{https://tether.to/wp-content/uploads/2021/05/tether-march-31-2021-reserves-breakdown.pdf}{official statement} by Tether Limited on 31 March 2021.} it is definite that both tether and Binance require more attention from academics, practitioners and especially regulators. 
Our conclusions should be useful  to inform the coming congress debates on the Digital Asset Market Structure and Investor Protection Act that is proposing sweeping reforms in crypto trading.


In terms of future research, we are interested to assess information processing and volatility transmission from the different perspective of volatility discovery. Using the methodology proposed in \citet{Dias2016} and applied to three spot exchanges by \citet{Dimpfl2021}, one could examine how much the different crypto instruments contribute to the common latent volatility process and whether the tether-margined perpetual contract on Binance -- which we identify as the main source of volatility -- is also the leading market in the volatility process. Given the declining importance of bitcoin within the crypto market, another interesting topic is a similar volatility transmission analysis within ether spot and perpetuals on different exchanges, or for ripple and other important coins -- and another analysis between bitcoin and ether and other coins, including both spot and perpetual instruments.

\singlespacing
\bibliographystyle{chicago}
\bibliography{VolaSpillover_Lit}

\end{document}